\documentclass[twocolumn,tighten,times]{aastex62}

\usepackage{amsmath}
\usepackage{hyperref} 

\received{January 1, 2018}
\revised{January 7, 2018}
\accepted{\today}
\submitjournal{ApJ}

\shorttitle{RRL and [NII] Observations in the Galactic Plane}
\shortauthors{Pineda, et  al.}

\begin{document}

\title{ Electron Densities and Nitrogen Abundances in Ionized Gas
  Derived using [N\,{\sc ii}] Fine--structure and Hydrogen
  Recombination lines.}

\correspondingauthor{Jorge L. Pineda}
\email{Jorge.Pineda@jpl.nasa.gov}

\author[0000-0001-8898-2800]{Jorge L. Pineda}
\affil{Jet Propulsion Laboratory, California Institute of  Technology, 4800 Oak Grove Drive, Pasadena, CA 91109-8099, USA}

\author{Shinji Horiuchi}
\affil{CSIRO Astronomy \& Space Science/NASA Canberra Deep Space Communication Complex, PO Box 1035, Tuggeranong ACT 2901, Australia}
\author[0000-0002-7045-9277]{Loren D. Anderson}
\affil{Department of Physics and Astronomy, West Virginia University, Morgantown, WV 26506, USA}
\affil{ Center for Gravitational Waves and Cosmology, West Virginia University, Chestnut Ridge Research Building, Morgantown, WV 26505, USA}
\affil{ Green Bank Observatory, P.O. Box 2, Green Bank, WV 24944, USA }
\author[0000-0001-8061-216X]{Matteo Luisi}
\affil{Department of Physics and Astronomy, West Virginia University, Morgantown, WV 26506, USA}
\affil{ Center for Gravitational Waves and Cosmology, West Virginia University, Chestnut Ridge Research Building, Morgantown, WV 26505, USA}
\author{William D. Langer}
\affil{Jet Propulsion Laboratory, California Institute of  Technology, 4800 Oak Grove Drive, Pasadena, CA 91109-8099, USA}
\author[0000-0002-6622-8396]{Paul F. Goldsmith}
\affil{Jet Propulsion Laboratory, California Institute of  Technology, 4800 Oak Grove Drive, Pasadena, CA 91109-8099, USA}
\author[0000-0003-1798-4918]{Thomas B. H. Kuiper}
\affil{Jet Propulsion Laboratory, California Institute of  Technology, 4800 Oak Grove Drive, Pasadena, CA 91109-8099, USA}
\author{Geoff Bryden}
\affil{Jet Propulsion Laboratory, California Institute of  Technology, 4800 Oak Grove Drive, Pasadena, CA 91109-8099, USA}
\author{Melissa Soriano}
\affil{Jet Propulsion Laboratory, California Institute of  Technology, 4800 Oak Grove Drive, Pasadena, CA 91109-8099, USA}
\author{T. Joseph W. Lazio}
\affil{Jet Propulsion Laboratory, California Institute of  Technology, 4800 Oak Grove Drive, Pasadena, CA 91109-8099, USA}



\begin{abstract}
  We present a method for deriving the electron density of ionized gas
  using the ratio of the intensity of the [N\,{\sc ii}] 205$\mu$m line
  to that of Hydrogen radio recombination lines (RRL). We use this
  method to derive electron densities of 21 velocity components in 11
  lines of sight through the Galaxy, including the Galactic center. We
  observed, at high--spectral resolution, the [N\,{\sc ii}] 205$\mu$m
  with the {\it Herschel}/HIFI and SOFIA/GREAT instruments and the
  radio recombination lines with the Green Bank Telescope and the NASA
  Deep Space Network Deep Space Station 43 (DSS--43) telescope. We
  find typical electron densities between 8 to 170\,cm$^{-3}$, which
  are consistent with those derived at low spectral resolution using
  the [N\,{\sc ii}] 205$\mu$m/122$\mu$m ratio with {\it Herschel}/PACS
  on a larger sample of sight lines in the Galactic plane. By matching
  the electron densities derived from the [N\,{\sc ii}] 205$\mu$m/RRL
  intensity ratio and the [N\,{\sc ii}] 122$\mu$m/205$\mu$m intensity
  ratio, we derive the nitrogen fractional abundance for most of the
  velocity components. We investigate the dependence of the N/H ratio
  with Galactocentric distance in the inner Galaxy ($R_{\rm
    gal}<$6\,kpc), which is inaccessible in optical studies due to
  dust extinction.  We find that the distribution of nitrogen
  abundances in the inner galaxy derived from our data has a slope
  that is consistent to that found in the outer Galaxy in optical
  studies. This result is inconsistent with some suggestions of a
  flatter distribution of the nitrogen abundance in the inner
  galaxy.  
\end{abstract}



\section{Introduction}
\label{sec:introduction}
The ionized gas component of the interstellar medium (ISM) occupies a large
volume in galaxies \citep{Haffner2009}. It is found in diffuse form as
the warm ionized medium (WIM), and in denser forms in H\,{\sc ii}
regions surrounding massive stars and in the ionized boundary layers
of molecular clouds. The structure and kinematics of the ionized gas
is a reflection of the radiative and mechanical feedback from massive
stars. Stellar feedback has been suggested to play a fundamental role
in the regulation of star formation in galaxies
\citep[e.g.][]{Hopkins2014}, which in turn drives galaxy
evolution. Thus, the study of the structure and kinematics of the
ionized gas is an important tool for characterizing the effect that
stellar feedback has in the ISM and for determining
the role of stellar feedback in the regulation of star formation in
galaxies.

After massive stars form, they produce energetic photons that ionize
their dense surroundings.  As the ionized gas is not in pressure
equilibrium with its surrounding neutral gas, H\,{\sc ii} regions
expand with time. Because the ionizing photon flux of a star is
relatively constant over its lifetime on the main sequence, the mass
of gas that it ionizes is relatively constant. Thus, as an H\,{\sc ii}
region evolves, it become larger and its volume density diminishes.
Eventually, the neutral gas surrounding H\,{\sc ii} regions becomes
porous, and ionizing photons escape into the diffuse ISM, as suggested
by power law decreases in RRL intensities surrounding H\,{\sc ii}
regions \citep{Luisi2016,Luisi2019}. This ionizing photon leakage,
which is more prominent in giant H\,{\sc ii} regions, can ionize
hydrogen over a larger volume surrounding H\,{\sc ii} regions and can
play an important role in maintaining the diffuse WIM
\citep{Haffner2009}.  The evolution of the ionized gas in star forming
regions is thus characterized by the volume density of electrons
($n_e$), which varies by orders of magnitude during the lifetimes of
H\,{\sc ii} regions, ranging from $> 10^4\,{\rm cm}^{-3}$ in
ultra--compact H\,{\sc ii} regions \citep{Churchwell2002} to
$\lesssim0.1\,{\rm cm}^{-3}$ \citep{Haffner2009,Cordes2002} in the
diffuse WIM. Mechanical feedback from massive stars, such as stellar
winds and supernova explosions, will further influence the density
structure of the ionized gas, creating features such as bubbles, shock
fronts, filaments, pillars, globules, and clumps.  Studying the
density structure and kinematics of ionized gas over large areas is
crucial for understanding the evolution of star-forming regions and
the impact of star formation on the ISM of galaxies.

Nitrogen, the fifth most abundant element, has an ionization potential
of 14.6 eV and therefore emission from ionized nitrogen arises
exclusively from the ionized gas component of the ISM. The
far-infrared [N\,{\sc ii}] 205$\mu$m and 122$\mu$m fine structure
lines are therefore important tracers of the highly ionized
low-density WIM, the extended envelopes of H\,{\sc ii} regions,
high-density H\,{\sc ii} regions, and partially ionized boundary
layers of photon dominated regions (PDRs). Emission from far-infrared
[N\,{\sc ii}] is widespread throughout the Galaxy as shown by COBE
\citep{Bennett1994}.  The far--infrared [N\,{\sc ii}] lines are among
the most important tracers of Galactic ionized gas as they are not
affected by dust extinction, which greatly restricts the variety of
environments that ionized gas tracers in the optical and near-infrared
can explore. 

\citet{Goldsmith2015} observed the [N\,{\sc ii}] 122$\mu$m and
205$\mu$m fine structure lines along $\sim$100 lines--of--sight using
the PACS instrument on {\it Herschel}, which had insufficient spectral
resolution to resolve the lines, and so they could only derive
integrated intensities.  The excitation analysis of the [N\,{\sc ii}]
122$\mu$m and 205$\mu$m lines indicates that the emission, over the
whole inner galaxy, arises from regions with relatively large volume
densities ($n_e\simeq$10--100 cm$^{-3}$), larger than those expected
for the diffuse WIM ($\lesssim$0.1\,cm$^{-3}$) but lower than those
typical of compact H\,{\sc ii} regions \citep[$>5\times10^3\,{\rm
  cm}^{-3}$;][]{Kurtz2005}.  The frequency and extent of the density
components found by \citet{Goldsmith2015} is inconsistent with them
arising only from compact H\,{\sc ii} regions, suggesting the presence
of an ubiquitous moderately dense ionized gas component in the plane
of the Milky Way \citep{Geyer2018}.  The {\it Herschel}/PACS
observations lack velocity information, and therefore, the resulting
electron densities could arise from multiple H\,{\sc ii} regions along
the line--of--sight, and thus the information of the distribution of
ionized gas in the Galactic plane is incomplete in this data set.
High--spectral resolution observations of the [N\,{\sc ii}] lines are
necessary to separate spatially overlapping gas components along the
line of sight and trace the kinematics of ionized gas. The HIFI
instrument on {\it Herschel} and the GREAT instrument on SOFIA, have
recently enabled the observations of the [N\,{\sc ii}] 205\,$\mu$m
line at high spectral resolution
\citep{Persson2014,Langer2016,Langer2017a}. However, neither of these
instruments could observe the [N\,{\sc ii}] 122\,$\mu$m line for use
with the [N\,{\sc ii}] 205$\mu$m line in the excitation analysis.

Radio recombination lines (RRLs) are also tracers of the ionized gas
observed with the [N\,{\sc ii}] 122\,$\mu$m and 205\,$\mu$m
lines. They have been extensively used to characterize the properties
of ionized gas regions \citep{Brown1978}. Early observations of
hydrogen recombination lines were limited to bright and dense H\,{\sc
  ii} regions due to the low sensitivity of radio telescopes at that
time.  Recent advances in radio spectrometer technology, however, have
enabled efficient observations of RRLs by observing a large number of
distinct transitions simultaneously, thus obtaining sensitive
observations by stacking several RRL lines. These new techniques have
allowed the study of RRL emission over a wider range of environments
and the detection of satellite recombination lines such as those from
helium and carbon \citep{Balser2006,Anderson2011,Alves2015,Luisi2017}.
The RRL lines provide an unambiguous determination of the emission
measure ($\int n_{\rm e}N({\rm H^+}) dl$; where $n_{\rm e}$ and
$N({\rm H^+})$ are the electron density and ionized hydrogen column
density, respectively) which can be combined with [N\,{\sc ii}]
205\,$\mu$m surveys done with {\it Herschel} and SOFIA to determine
the electron density of ionized gas at high velocity resolution.
[N\,{\sc ii}] and RRL lines together can be used to provide a more
complete picture of the properties of ionized gas in star forming
regions.

In this paper, we present a new approach that uses spectrally resolved
hydrogen recombination lines together with the [N\,{\sc ii}] 205$\mu$m
line to determine the electron density. We apply this method in 11
lines--of--sights distributed in the Galactic plane observed with {\it
  Herschel}/HIFI and SOFIA in [N\,{\sc ii}] 205$\mu$m and the DSS--43
and GBT telescopes in Hydrogen RRLs.  This paper is organized as
follows. We describe the RRL and [N\,{\sc ii}] observations in our
selected sample in Section~\ref{sec:observations}. In
Section\,\ref{sec:determ-electr-dens}, we discuss the method used to
derive electron densities from the ratio of the RRL to the [N\,{\sc
  ii}] 205$\mu$m lines. In Section\,\ref{sec:discussion-2} we discuss
the derived electron densities in the selected LOSs. We finally list
our conclusions in Section~\ref{sec:conclusions}.

\section{Observations}
\label{sec:observations}

\subsection{[N\,{\sc ii}] Observations}
\label{sec:n-sc-ii-1}
  
We observed 10 lines--of--sights (LOSs) in the Galactic plane in the
[N\,{\sc ii}] 205\,$\mu$m line at high spectral resolution using the
HIFI \citep{deGraauw2010} instrument aboard the {\it Herschel Space
  Observatory} \citep{Pilbratt2010}. These LOSs are part of the
GOT\,C+ survey \citep{Langer2010,Pineda2013} and the [N\,{\sc ii}]
observations were presented by \citet{Goldsmith2015} and
\citet{Langer2016}.  The intensities were converted from antenna
temperature ($T^*_{\rm A}$) scale, to main beam brightness
temperature, $T_{\rm mb}$, by using a main beam efficiency $\eta_{\rm
  mb}$ of 0.60 \citep[][updates in HIFI Beam Release Notes 2014
October]{Roelfsema2012}. The data were smoothed to a velocity
resolution of 1\,km\,s$^{-1}$ in order to increase the signal-to-noise
ratio.  The FWHM width of the {\it Herschel}/HIFI beam corresponds to
15.7\arcsec\ for the [N\,{\sc ii}] 205$\mu$m line.  The typical rms antenna
temperature of this data set is $T_{\rm mb}$=0.1\,K in a
1\,km\,s$^{-1}$ channel. For the diffraction-limited {\it Herschel}
HIFI beam, this rms is equivalent to an uncertainty in the intensity of
$\Delta I = 3\times10^{-9}$ W\,m$^{-2}$\,sr$^{-1}$.

We also included in our analysis the GOT\,C+ LOS, G030.0+0.0, that was
observed with SOFIA/GREAT and presented by \citet{Langer2017}. The
angular resolution of these observations is 20\arcsec. We converted
from $T^*_{\rm A}$ to $T_{\rm mb}$ intensity scale using a main beam
efficiency of 0.66 \citep{Roellig2016}. The spectrum has an rms noise
of 0.12\,K in a 1 km\,s$^{-1}$ channel.

The observed {\it Herschel} and SOFIA [C\,{\sc ii}] and [N\,{\sc ii}]
spectra together with RRL observations described below are shown in
Figure~\ref{fig:spectra}. We also list their integrated intensities in
Table~\ref{tab:intensities} and the results of a Gaussian
decomposition in Table\,\ref{tab:gauss-decomp-1}. There is general
agreement among the velocities of [C\,{\sc ii}], [N\,{\sc ii}], and
RRL components.  There are, however, differences in the [N\,{\sc
    ii}]/RRL line ratio, which is a signature of different physical
conditions along the line--of--sight (see Section~\ref{sec:results}).
These differences in the [N\,{\sc ii}]/RRL ratio in LOSs with multiple
velocity components in a small velocity range, mostly seen in the
intrinsically narrower [N\,{\sc ii}] lines, result in apparent
velocity shifts between [N\,{\sc ii}] and RRL lines
(e.g. G013.9+0.0). However, when decomposing the data in Gaussian
components, we find a good correspondence between the [N\,{\sc ii}]
and RRL velocity components even in these LOSs, confirming the
assumption that they both trace the same ionized gas.

\subsection{Hydrogen Recombination Lines}
\label{sec:hydr-recomb-lines}

We used the DSS--43 telescope to observe 5 LOSs in our sample in the
H91$\alpha$ and H92$\alpha$ hydrogen radio recombination lines at
8.58482\,GHz and 8.30938\,GHz respectively, using the X--band receiver
in the position--switching observing mode. The angular resolution of
the DSS--43 at 8.420\,GHz is 115.2\arcsec. We converted the data from
an antenna temperature to a main beam temperature scale using a main
beam efficiency of 0.78.  Both lines were resampled to a common
spectral grid and averaged together to increase the signal--to--noise
ratio, with the H92$\alpha$ intensities scaled to correspond to that
of the H91$\alpha$ lines\footnote{As seen in Equation~(\ref{eq:3}), in
  LTE, the intensity of a RRL is proportional to the $EM$, the line
  width, and line frequency. Given that $EM$ and $\Delta v$ are
  intrinsic properties of the source, the intensity of two RRLs are
  related by the inverse of the ratio of their frequencies. In the
  case when LTE does not apply (Equation~\ref{eq:7}), however, a
  dependence on electron density and temperature is introduced to the
  relationship between two line intensities. For lines with similar
  principal quantum number, $n$, this effect is negligible, and
  assuming LTE is appropriate. For larger $\Delta n$, however, NLTE
  effects need to be taken into account when scaling RRL intensities.
} .  We fitted a polynomial baseline of order 3 to our data. The
resulting spectra has a typical rms noise of 4\,mK in a
1\,km\,s$^{-1}$ channel.  We also used the DSS--43 telescope to
observe the H67$\alpha$ (21.3846\,GHz) at 47\arcsec\ in the G345.7+0.0
LOS. We converted the spectrum from an antenna temperature to a main
beam temperature scale using a main beam efficiency of 0.5 and we
fitted a polynomial baseline or order 3. The spectra have a rms noise
of 6\,mK over 3\,km\,s$^{-1}$ channel.

We also observed RRLs in 6 LOSs in our sample using the Versatile GBT
Astronomical Spectrometer (VEGAS) on the Green Bank Telescope (GBT) in
X--band in the position--switching observing mode.  The angular
resolution of the GBT in X--band is 84\arcsec. For each observed
direction, we simultaneously measured seven H$n\alpha$ RRL transitions
in the 9 GHz band, H87$\alpha$ to H93$\alpha$, using the techniques
discussed in \citet{Bania2010}, \citet{Anderson2011}, and
\citet{Balser2011}, and averaged all spectra together to increase the
signal-to-noise ratio using {\tt TMBIDL} \citep{Bania2016}. The data
were resampled to a common grid, intensities scaled to correspond to
that for the H89$\alpha$ line (9.17332\,GHz), and we averaged all
lines in the band (2 polarizations per transition) together to
increase the signal--to--noise ratio. The GBT data was calibrated
using a noise diode of known power.  The data were later corrected
with a 3rd order polynomial baseline and smoothed to
$\sim$1.9\,km\,s$^{-1}$. We converted the intensities from an antenna
temperature to main beam temperature using a main beam efficiency of
0.94. The typical rms noise of this data is 2.5\,mK in a 1.9
km\,s$^{-1}$ channel.

We also observed the H53$\alpha$ (42.95196\,GHz) and H54$\alpha$
(40.63049\,GHz) lines with the Q--Band receiver on the GBT in the
G000.0+0.0 LOS. These observations have an angular resolution of
16\arcsec\ which is similar to that of the {\it Herschel} HIFI beam,
and can therefore be used to study beam dilution effects. The spectra
were averaged to increase the signal--to--noise ratio, with the
H54$\alpha$ used as a reference.  We converted the data from an
antenna temperature to a main beam temperature scale using a main beam
efficiency of 0.8 and we fitted a polynomial baseline of order 3. The
spectra have a rms noise of 5\,mK over 1.1\,km\,s$^{-1}$ channel.

\begin{deluxetable*}{lrrcccccc} 
\tabletypesize{\footnotesize} \centering \tablecolumns{9} \small
\tablewidth{0pt} \tablecaption{[NII] 205$\mu$m and RRL integrated intensities$^{1}$} \tablenum{1}
\tablehead{\colhead{LOS} & \colhead{$l$} & \colhead{$b$} & \colhead{V$_{\rm LSR}$} & \colhead{$I({\rm [NII]})$ 205$\mu$m} & \colhead{ $I({\rm H91\alpha})$   } & \colhead{$I({\rm H89\alpha})$} & \colhead{$I({\rm H67\alpha})$} & \colhead{$I({\rm H54\alpha})$}  \\ 
\colhead{}   & \colhead{[\degr]}   & \colhead{[\degr]}   & \colhead{[km\,s$^{-1}$]} & \colhead{[K\,km\,s$^{-1}$]} &   \colhead{[K\,km\,s$^{-1}$]} & \colhead{[K\,km\,s$^{-1}$]} &   \colhead{[K\,km\,s$^{-1}$]} & \colhead{[K\,km\,s$^{-1}$]} \\}
\startdata 
G305.1+0.0 & 305.106 & 0.0 & -33.2 &	37.6$\pm$0.9  &  1.58$\pm$0.027$^{2}$  & -- &  -- & --   \\  
G316.6+0.0 & 316.596 & 0.0 & -48.1 &	20.9$\pm$0.4  &  0.57$\pm$0.027$^{2}$  & -- &  -- & --   \\  
G316.6+0.0 & 316.596 & 0.0 & -6.4 &	3.4$\pm$0.4    &  0.12$\pm$0.013$^{2}$  & -- &  -- & --   \\  
G342.2+0.0 & 342.174 & 0.0 & -131.2 &	9.7$\pm$0.6   &  0.30$\pm$0.021$^{2}$  & -- &  -- & --   \\  
G337.0+0.0 & 336.957 & 0.0 & -121.5 &	25.1$\pm$0.5   &  1.13$\pm$0.064$^{2}$  & -- &  -- & --   \\  
G337.0+0.0 & 336.957 & 0.0 & -76.6 &	21.3$\pm$2.3   &  1.31$\pm$0.080$^{2}$  & -- &  -- & --   \\  
G345.7+0.0 & 345.652 & 0.0 & -122.8 &	6.2$\pm$0.4   &  0.35$\pm$0.087$^{2}$  &  --  & --    & --   \\  
G345.7+0.0 & 345.652 & 0.0 & -8.2 &	16.2$\pm$0.6  &  2.18$\pm$0.010  & --  & 0.89$\pm$0.092$^{2}$ &  --   \\  
G349.1+0.0 & 349.130 & 0.0 & 17.0 &	23.0$\pm$0.8  &  2.25$\pm$0.069 & 3.26$\pm$0.014$^{2}$ & --  &  --  \\  
G349.1+0.0 & 349.130 & 0.0 & -91.1 &	11.5$\pm$0.9  &  0.73$\pm$0.076 & 0.66$\pm$0.007$^{2}$ & --  &  --  \\  
G000.0+0.0 & 0.000 & 0.0 & -60.5 &	33.2$\pm$1.7  &  4.251$\pm$0.193 & 3.81$\pm$0.018$^{2}$ & --  & 0.744$\pm$0.073  \\  
G000.0+0.0 & 0.000 & 0.0 & -37.1 &	45.7$\pm$2.5  &  2.173$\pm$0.204 & 2.40$\pm$0.028$^{2}$ & --  & --  \\  
G000.0+0.0 & 0.000 & 0.0 & 12.8 &	13.2$\pm$2.3  &  0.516$\pm$0.154 & 1.33$\pm$0.019$^{2}$ & --  & --  \\  
G000.0+0.0 & 0.000 & 0.0 & 95.0 &	14.8$\pm$0.2  &  0.867$\pm$0.118 & 0.70$\pm$0.008$^{2}$ & --  & --   \\  
G013.9+0.0 & 13.913 & 0.0 & 45.5 &	2.3$\pm$0.3   &  --              & 0.14$\pm$0.006$^{2}$ & --  & --  \\  
G013.9+0.0 & 13.913 & 0.0 & 30.3 &	1.5$\pm$0.3   &  0.373$\pm$0.075 & 0.49$\pm$0.006$^{2}$ & --  &  --  \\  
G030.0+0.0 & 30.000 & 0.0 & 95.4 &	10.2$\pm$0.7  &  --              & 0.38$\pm$0.030$^{2}$ & --  &               --  \\  
G031.3+0.0 & 31.277 & 0.0 & 100.4 &	20.9$\pm$1.4   &  0.678$\pm$0.197 & 0.63$\pm$0.005$^{2}$ & --  & --  \\   
G031.3+0.0 & 31.277 & 0.0 & 38.0 &	2.8$\pm$0.9   &  -- & 0.51$\pm$0.007$^{2}$ & --  &               --  \\  
G049.1+0.0 & 49.149 & 0.0 & 59.7 &	3.3$\pm$0.3   &  0.078$\pm$0.030 & 0.13$\pm$0.004$^{2}$ & --  &  --  \\  
\enddata
\tablenotetext{1}{The intensities shown here are not corrected for
  beam dilution effects (see Section\,\ref{sec:beam-dilut-effects}).}
\tablenotetext{2}{RRL intensities that are used in our analysis.}
\label{tab:intensities}
\end{deluxetable*} 

\subsection{Sample location with respect to known H\,{\sc ii} Regions}
\label{sec:sample-location-with}

\begin{figure*}[t]
\centering
\includegraphics[width=0.8\textwidth,angle=0]{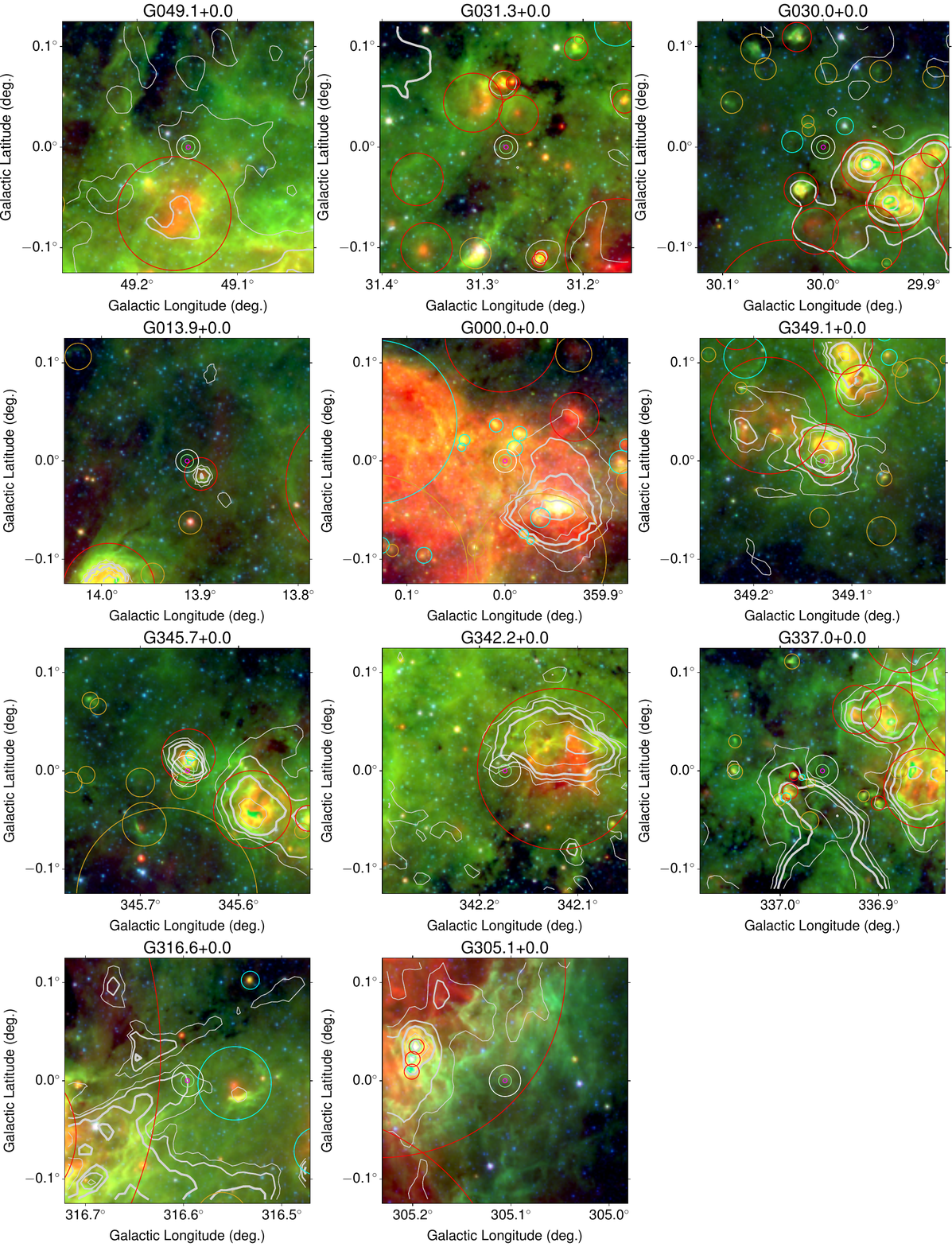}
\caption{Location of our LOS sample overlaid in {\it Spitzer}
  24$\mu$m MIPS in red, GLIMPSE 8$\mu$m in green, and GLIMPSE
  3.6$\mu$m in blue.  At the center of these images, the white circle
  corresponds to the beam of the DSS--43 or GBT RRL observations, the
  magenta circle that of the {\it Herschel}/HIFI [N\,{\sc ii}]
  205$\mu$m observations, and the green circle denotes approximately
  the size of the {\it Herschel}/PACS 122$\mu$m and 205$\mu$m
  footprint.  Contours are SUMSS 843\,MHz for LOSs with
  $305.1\degr\leq l \leq 0\degr$, NVSS 1.4\,GHz for G013.9+0.0, and VGPS 1.4\,GHz
  for sources with $30.0\degr \leq l \leq 49.1\degr$, set to 5, 10, 20, 40, and 80\% of
  the peak.  Circles are red for known H\,{\sc ii} regions, cyan for
  candidates that have detected radio continuum, and orange for
  candidates that do not have radio continuum.
}
\label{fig:environment}
\end{figure*}

We studied the environment traced by our sample LOS by searching for
the nearest known H\,{\sc ii} regions from the Wide-field Infrared
Survey Explorer (WISE) Catalog of Galactic H\,{\sc ii} Regions
\citep{Anderson2014}. These regions were followed up by RRL and radio
continuum observations to confirm that the mid--infrared warm dust
emission is associated with ionized gas
\citep{Bania2010,Anderson2011,Wenger2019}. In Table~\ref{tab:xmatch3},
we list the nearest WISE H\,{\sc ii} region to each LOS in our sample,
the distance between the center of the observations' beam and that of
the nearest H\,{\sc ii} region, and the radius of the nearest H\,{\sc
  ii} region. The radius of a WISE H\,{\sc ii} region is defined by
that of a circular aperture that encloses its associated mid--infrared
emission.  In Figure~\ref{fig:environment}, we show the location of
the RRL and [N\,{\sc ii}] beam and also of nearby sources in the WISE
H\,{\sc ii} region catalog overlaid in {\it Spitzer} 24$\mu$m MIPS,
GLIMPSE 8$\mu$m, GLIMPSE 3.6$\mu$m images tracing dust continuum
emission. We also include contour lines showing the distribution of
radio continuum emission around these sources. We find that most of
our LOSs are located in the vicinity of H\,{\sc ii} regions but do not
overlap with their brightest parts, and thus are unlikely to be
associated with compact H\,{\sc ii} regions.
%
%
However, there are two exceptions, G345.7+0.0 and G349.1+0.0 in which
the beam of our observations partially overlap with the brightest
parts of H\,{\sc ii} regions.

\begin{figure*}[t]
\centering
\includegraphics[width=0.6\textwidth,angle=0]{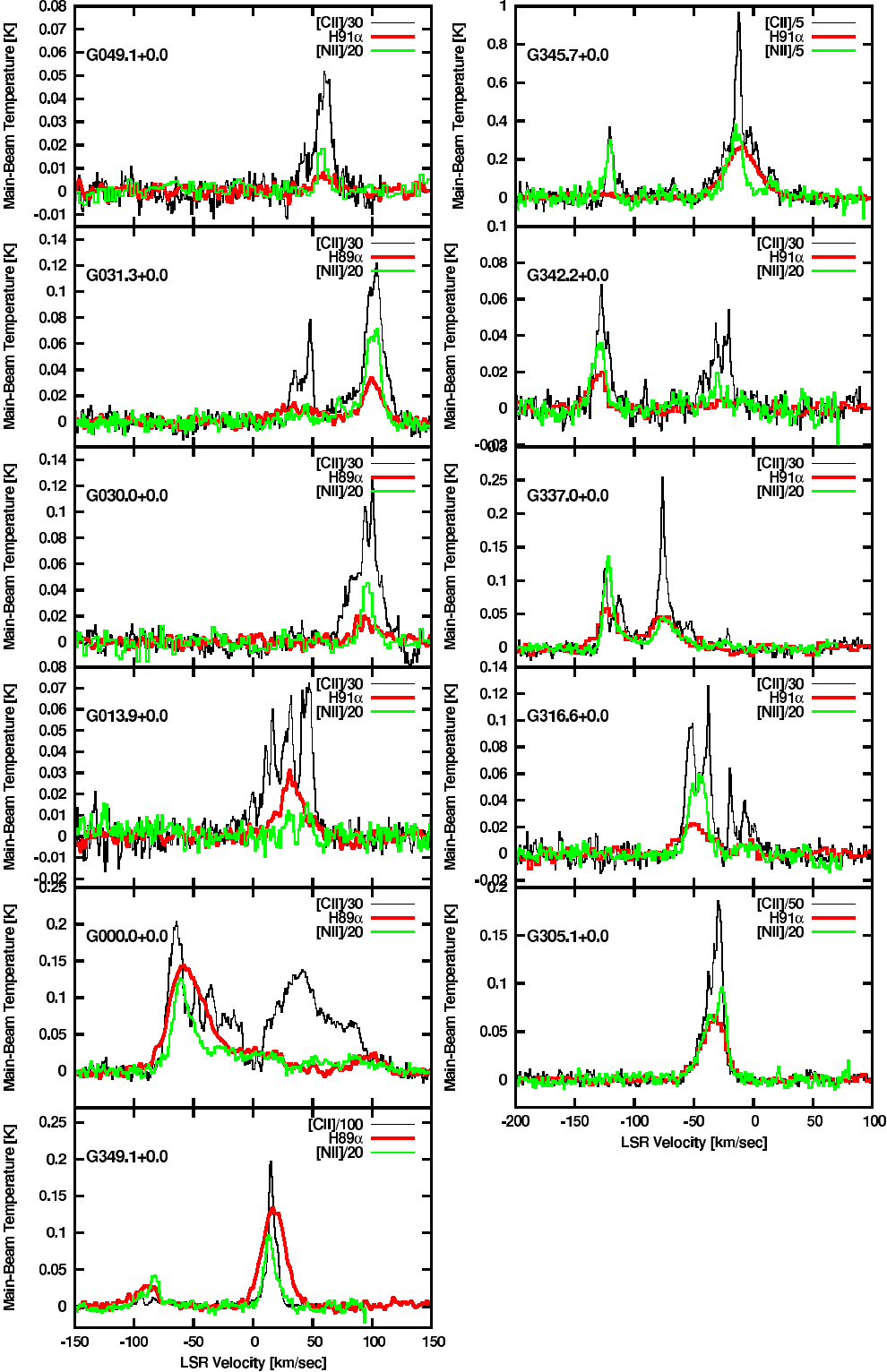}
\caption{Observed spectra of [C\,{\sc ii}], [N\,{\sc ii}], and
  Hydrogen recombination lines along the 11 LOSs in our sample. }\label{fig:spectra}
\end{figure*}

\begin{deluxetable}{lcccccccc} 
\tabletypesize{\footnotesize} \centering \tablecolumns{4} \small
\tablewidth{0pt} \tablecaption{Nearest known H\,{\sc ii} region to sample LOSs} \tablenum{2}
\tablehead{\colhead{LOS} &  \colhead{WISE H\,{\sc ii} Source} & \colhead{Distance} & \colhead{WISE H\,{\sc ii} Radius}  \\ 
\colhead{}  & \colhead{} &   \colhead{\arcsec} &    \colhead{\arcsec}  \\}
\startdata
G305.1+0.0 & G305.201+00.009  &343.6 & 27.6  \\
G316.6+0.0 & G316.548-00.003  &173.1 & 134.5  \\
G337.0+0.0 & G336.969-00.013  &61.7  & 17.3  \\
G345.7+0.0 & G345.651+00.015  &55.5  & 98.2  \\
G342.2+0.0 & G342.120+00.001  &194.5 & 295.2  \\
G349.1+0.0 & G349.126+00.010  &40.7  & 98.9  \\
G000.0+0.0 & G000.008+00.036  &134.9 & 25.0  \\ 
G013.9+0.0 & G013.899-00.014  &69.3  & 60.0  \\
G030.0+0.0 & G030.014+00.017  &80.5  & 23.7  \\
G031.3+0.0 & G031.264+00.031  &123.4 & 71.0  \\
G049.1+0.0 & G049.163-00.066  &243.0 & 203.6  \\
\enddata
\tablenotetext{}{}
\label{tab:xmatch3}
\end{deluxetable}

\section{Determination of the electron density with [N\,{\sc ii}] and RRL observations}
\label{sec:determ-electr-dens}
\subsection{ [N\,{\sc ii}] 205 $\mu$m Emission}
\label{sec:n-sc-ii}

\begin{figure}[t]
\centering
\includegraphics[width=0.45\textwidth,angle=0]{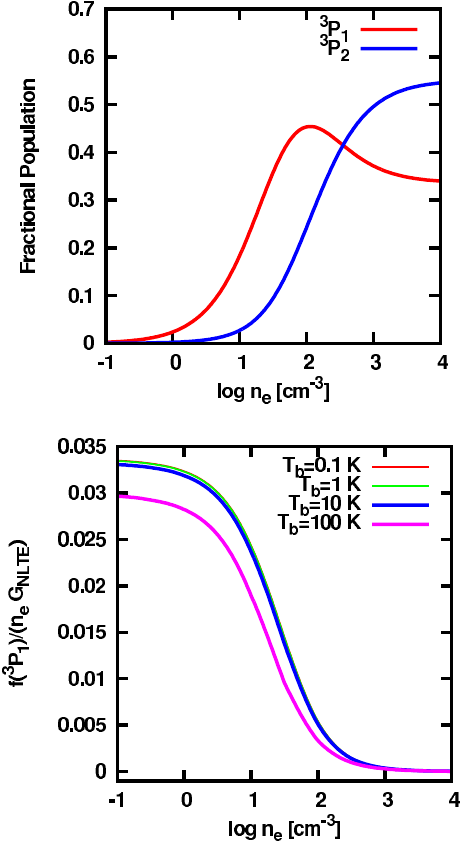}
\caption{({\it upper panel}) The fractional population in the $^3{\rm
    P}_1$ and $^3{\rm P}_2$ levels of ionized nitrogen as a function
  of the electron density for $T_e$=8000\,K. ({\it bottom panel}) The
  dependence of the [N\,{\sc ii}] 205$\mu$m/RRL ratio on electron
  density for a set of continuum brightness temperatures (see
  Equation~\ref{eq:9}).  }\label{fig:fnu_dependence}
\end{figure}

The ground electronic state of ionized nitrogen is a three fine
structure level system which results in two fine-structure transitions
at 122$\mu$m ($^3$P$_2$-$^3$P$_1$) and 205$\mu$m
($^3$P$_1$-$^3$P$_0$). The brightness of a [N\,{\sc ii}] line
integrated over frequency for an optically thin spectral line of
frequency $\nu_{ul}$, spontaneous decay rate, $A_{ul}$, and upper
level column density $N_u$, is given by \citep[e.g.][]{Goldsmith2012},

\begin{equation}\label{eq:1}
I=\frac{A_{ul} h \nu_{ul} N_u }{4 \pi}\,{[\rm erg\,cm^{-2}\,s^{-1}\,sr^{-1}]},
\end{equation}
or in terms of the main beam temperature per unit velocity,

\begin{equation}\label{eq:2}
\int T_{\rm mb}d{ v}=\frac{c^3}{2k_{\rm b} \nu_{ul}^3} I=\frac{A_{ul} h  c^3 N_u }{8  \pi k_{b} \nu^2_{ul} }{\rm [K\,km\,s^{-1}]},
\end{equation}
For the [N\,{\sc ii}] 205$\mu$m line, we can write the column density
at the upper level (${\rm ^3P_1}$) in terms of the total N$^+$ column
density as, $N({\rm ^3P_1})=f({\rm ^3P_1})N(\rm N^+)$, where $f({\rm
  ^3P_1})$ is the fractional population of the $^3{\rm P}_1$ level. We
can therefore write Equation~(\ref{eq:2}) as,
\begin{equation}
\begin{split}
\int T^{[\rm NII]}_{\rm mb}{dv}= & \frac{A_{10} h  c^3 }{8  \pi k \nu_{10}^2 }f({\rm ^3P_1}) N({\rm N^+})\\
 & =5.01\times10^{-16}f({\rm ^3P_1}) N({\rm N^+})\,{\rm [K\,km\,s^{-1}]}.
\end{split}
\label{eq:8}
\end{equation}
For this transition, the spontaneous decay rate (Einstein's $A$
coefficient) is $A_{10}=2.08\times10^{-6}$\,s$^{-1}$, and the rest
frequency is $\nu_{10}=1.461\times10^{12}$\,Hz. \citet{Goldsmith2015}
calculated the fractional population of the $^3{\rm P}_1$ and $^3{\rm
  P}_1$ levels as a function of the electron density of ionized gas,
as shown in the upper panel of Figure~\ref{fig:fnu_dependence}.

\begin{figure}[t]
\centering
\includegraphics[width=0.45\textwidth,angle=0]{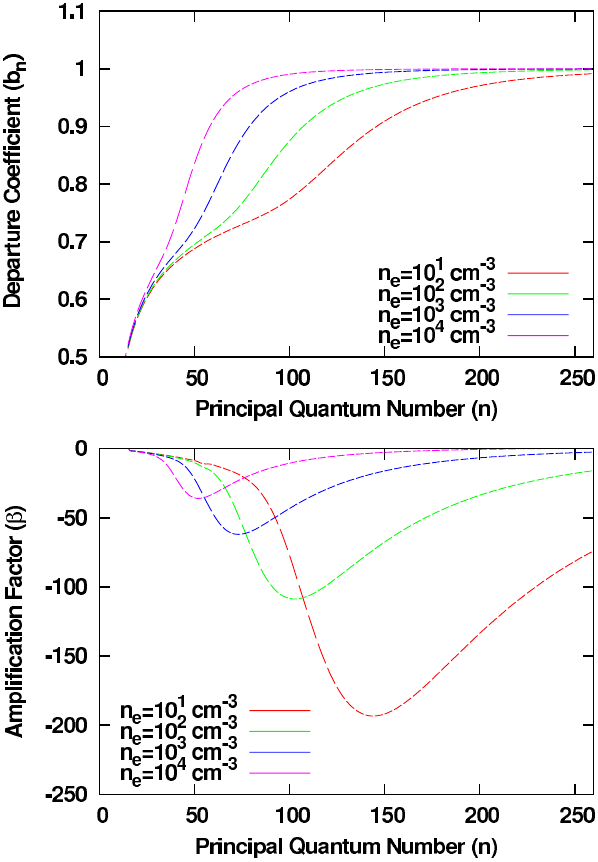}
\caption{The effects of deviations from local thermodynamical
  equilibrium are well understood \citep{Gordon2002} and a correction
  for these effects can be readily applied. Here we show the
  dependence of the departure coefficient ({\it upper panel}) and
  amplification factor ({\it lower panel}) as a function of principal
  quantum number for a range of electron
  densities.}\label{fig:gordon_comparion}
\end{figure}

\subsection{Hydrogen Recombination Line Emission}
\label{sec:hydr-recomb-line}

In local thermodynamical equilibrium (LTE), the main beam temperature
(in units of K) per unit velocity (km s$^{-1}$) of a hydrogen
recombination line is given by \citep{Tools},
\begin{equation}\label{eq:3}
\int T^{\rm RRL}dv=5.76\times10^{11} T_{\rm e}^{-3/2} EM \nu^{-1}, 
\end{equation}
where the line frequency $\nu$ is in Hz, the electron
temperature $T_e$ is in K, and the emission measure (EM) is in
cm$^{-6}$\,pc.  The EM is defined as the integral of the electron
volume density squared along the line of sight,
\begin{equation}\label{eq:4}
EM=\int n^2_e dl.
\end{equation}
The narrow range of [N\,{\sc ii}] and RRL emission as a function of
velocity seen in our data set (Figure~\ref{fig:spectra}) indicates it
arises from discrete sources rather than from the extended diffuse
warm ionized medium, which would show emission over a much wider range
of velocities. Thus, we can assume that the electron density, $n_e$,
is approximately constant along the line of sight, so that this
equation can be simplified to
\begin{equation}\label{eq:5}
EM=n_eN_e\simeq n_eN({\rm H^+}),
\end{equation}
where $N_e$ and $N({\rm H^+}$) are the column densities of electrons
and ionized hydrogen, respectively. We can thus, write equation
(\ref{eq:3}) in terms of the H$^+$ column density and electron density
as
\begin{equation}\label{eq:6}
\int T^{\rm RRL}dv=1.87\times10^{-7} \nu_{RRL}^{-1} T_{\rm e}^{-3/2} n_{\rm e} N({\rm H }^+), 
\end{equation}
where $n_{\rm e}$ is in units of cm$^{-3}$ and $N({\rm H }^+)$ in units of cm$^{-2}$.

The hydrogen recombination line emission can be affected by deviations
from local thermodynamical equilibrium. This deviation can be
defined in terms of the ratio
\begin{equation}\label{eq:7}
G_{\rm NLTE}(n_e, T_c)=\frac{ T^{\rm RRL}}{T^{\rm RRL}_{\rm NLTE}}=b_n \left [1-\frac{1}{2}\tau_{\rm c} \beta_n  \right ],
\end{equation}
where $b_n$ and $\beta_n$ are the departure coefficient and
amplification factor, for a principal quantum number $n$,
respectively, and $T_c$ and $\tau_c$ are the continuum brightness
temperature and opacity, respectively, at $\nu_{RRL}$. In
Figure~\ref{fig:gordon_comparion} we show the departure coefficient
and amplification factor as a function of the principal quantum number
for a set of electron densities.  The continuum opacity can be derived
from observations of $T_c$, and assuming an electron temperature,
using $\tau_c=T_{\rm c}/T_{\rm e}$.  The effects of deviations from
local thermodynamical equilibrium are well understood
\citep{Gordon2002} and a correction for these effects can be readily
applied. We roughly estimated $T_{\rm c}$ in our sources by
interpolating or extrapolating the flux at the frequency of our RRL
observations using data at two other frequencies, assuming a
power--law spectral energy distribution.  For sources with $l <
0$\degr\ we used the Parkes $\lambda$6cm (5\,GHz;
\citealt{Haynes1978}) and SGPS $\lambda$21cm (1.4\, GHz;
\citealt{Haverkorn2006}) radio continuum, while for sources with $l >
0$\degr\ we used the Parkes 6cm (5\,GHz) and Nobeyama $\lambda$3cm
(10\,GHz; \citealt{Handa1987}) surveys. One exception was G049.1+0.0
where we used the Effelsberg $\lambda$11cm (2.7\,Ghz;
\citealt{Reich1984}) radio continuum map instead of the Parkes
$\lambda$6cm data, as this LOS is not available in the latter survey.
All maps were convolved to a common resolution of 4\arcmin. For the
Galactic center position, G000.0+0.0, we extracted the brightness
temperature using the GBT X-Band radio continuum map presented by
\citet{Law2008}. The resulting continuum temperatures are listed in
Table~\ref{tab:densities}.  We find that, except for G000.0+0.0 with
$T_c=$7\,K, all continuum temperatures in our sample are below 1\,K.
As shown in the lower panel of Figure~\ref{fig:fnu_dependence}, the
[N\,{\sc ii}] to RRL ratio, defined below, has a weak dependence on
the continuum brightness temperature, mainly influencing the low
density regime for $T_c>$10K. Thus, we expect that deviations from
local thermodynamic equilibrium have a negligible impact in our
derived electron densities. Even if the uncertainties in our derived
continuum temperatures is a factor 10, they would result in only a 3\%
difference in the derived electron densities.


By combining Equations~(\ref{eq:8}),~(\ref{eq:6}), and (\ref{eq:7}),
we find that the ratio of the [N\,{\sc ii}] to RRL lines, $R^{\rm
  [NII]}_{\rm RRL}$, is a function of the electron density, the
electron temperature, the continuum brightness temperature, and the
relative abundance of ionized nitrogen with respect to ionized
hydrogen, $X_{\rm N}=\frac{N({\rm N}^+)}{N({\rm H}^+)}$, and is given
by,

\begin{equation}\label{eq:9}
R^{\rm [NII]}_{\rm RRL}=2.68\times10^{-9}\nu_{\rm RRL}T_{\rm e}^{3/2}\frac{f({\rm ^3P_1})}{n_e}G^{-1}_{\rm NLTE}(n_e,T_c) X_{\rm N}.
\end{equation}

In the lower panel of Figure~\ref{fig:fnu_dependence} we show the
portion of Equation~(\ref{eq:9}) that directly depends on the electron
density, $f({\rm ^3P_1})/(n_eG_{\rm NLTE}(n_e,T_c))$, as a function of
electron density for several continuum brightness temperatures. The
[N\,{\sc ii}]/RRL line ratio is a sensitive probe of electron density
between 3 and 300\,cm$^{-3}$, which is a typical density range for
evolved H\,{\sc ii} regions in the Galactic plane
\citep{Goldsmith2015}.  For each of the velocity components studied
here, we assume an electron temperature (in units of K) from the fit
by \citet{Balser2015} as a function of Galactocentric distance (in
kpc),
\begin{equation}\label{eq:20}
T_e = (4446 \pm 301) + (467 \pm 34)R_{\rm gal}.
\end{equation}
We also assume a nitrogen abundance relative to hydrogen, $X_{\rm N}$,
from the fit as a function of Galactocentric distance (in kpc) presented by
\citet{Esteban2018},
\begin{equation}\label{eq:21}
  12+\log({\rm N}/{\rm H})=8.21(\pm0.09)-0.059(\pm0.009)R_{\rm gal}.
\end{equation}
The distance to the Galactic center $R_{\rm gal}$ for a given velocity
component with Galactic longitude $l$, latitude $b$, and local
standard of rest (LSR) velocity $V_{\rm LSR}$, is given by
\begin{equation}
\label{eq:11}
R_{\rm gal}= R_{\odot} \sin (l)\cos(b) \left (\frac{V(R_{\rm gal})}{V_{\rm LSR}+V_{\odot}\sin(l)\cos(b)} \right ),
\end{equation}
where $R_{\odot}$ is the distance from the Sun to the Galactic center
and $V_{\odot}$ is the orbital velocity of the Sun with respect to the
Galactic center, and $V(R_{\rm gal})$ is the rotation curve.  We
determined $R_{\rm gal}$ for a given $l$, $b$, and $V_{\rm LSR}$ using
Equation~\ref{eq:11} assuming the "Universal" rotation curve presented
by \citet{Persic1996} (see Equation 3 in \citet{Wenger2018}) with
parameters fitted using trigonometric parallaxes and proper motions
for masers associated with massive star formation presented by
\citet{Reid2014}. This fit results in values of $R_{\odot}$ =
8.31\,kpc and $V_{\odot}$ = 241\,km\,s$^{-1}$.  We assumed $R_{\rm
  gal}= 0$\,kpc for the velocity components observed towards the
Galactic center, as they coincide with the velocity components
observed in different molecules by \citet{Henshaw2016}, and that are
all interpreted to be part of the central molecular zone. In
Table~\ref{tab:densities}, we list the derived $R_{\rm gal}$, $T_{\rm
  e}$, and $X_{N}$ for each velocity component.

\subsection{Uncertainties}
\label{sec:uncertainties}

The main factors influencing the uncertainties in the derivation of
electron densities from the [N\,{\sc ii}]/RRL ratio are those arising
from the uncertainties in the assumed electron temperature, and the
assumed relative abundance of nitrogen with respect to hydrogen, and
the uncertainties in the measurements.  The typical scatter in the
electron temperature found by \citet{Quireza2006} is about $\pm$350\,K
and results in a typical 13\% uncertainty in the derived electron
density. The typical scatter from the fit to $X_{\rm N}$ as a function
of Galactocentric distance presented by \citet{Esteban2018}, is about
23\%, resulting in a typical 44\% uncertainty in the derived electron
density.  We estimated the uncertainty in the [N\,{\sc ii}]/RRL ratio
by propagating those of the integrated intensities of the [N\,{\sc
  ii}] and RRL lines.  We find typical uncertainties of 20\% in the
measurement of the [N\,{\sc ii}]/RRL ratio.  To estimate the total
error in our electron density measurements we added the contribution
from each source of uncertainty in quadrature. The total error for the
derived electron density is typically 49\%.

\subsection{Beam Dilution Effects}
\label{sec:beam-dilut-effects}

In our analysis we use data at different angular resolutions
(16\arcsec\ for [N\,{\sc ii}] and 84\arcsec\ or 115\arcsec\ for the
RRL data), and therefore our derived electron densities can be
affected by beam dilution effects.  In
Appendix~\ref{sec:beam-dilut-corr}, we evaluated the effects of beam
dilution in our sample using different methods. We applied a
correction factor to the 16\arcsec\ [N\,{\sc ii}] data using spatial
information provided by {\it Herschel}/PACS observations from
16\arcsec\ (single pixel) to 47\arcsec\ ($5\times5$ pixel average).  We
also corrected the RRL data for beam dilution effects between
47\arcsec, 84\arcsec, and 115\arcsec, using the spatial information
provided by high resolution radio continuum data.  We summarize the
dilution factors, and their impact on the derived electron densities,
in Table~\ref{tab:dilution}.  By using these dilution factors we find
that the derived electron densities would vary by $\sim$15\% for
sources observed in RRLs at 84\arcsec\ and by $\sim$36\% for sources
observed at 115\arcsec.  We also applied beam dilution corrections to
the data in the G000.0+0.0 and G345.7+0.0 LOSs derived by comparing
RRL data observed at different angular resolutions.  By applying these
dilution factors to our data we are estimating electron densities on
scales of 47\arcsec, except for G000.0+0.0 were densities are derived
on scales of 16\arcsec.

\section{Discussion}
\label{sec:discussion-2}

%

\begin{figure*}[t]
\centering
\includegraphics[width=0.8\textwidth,angle=0]{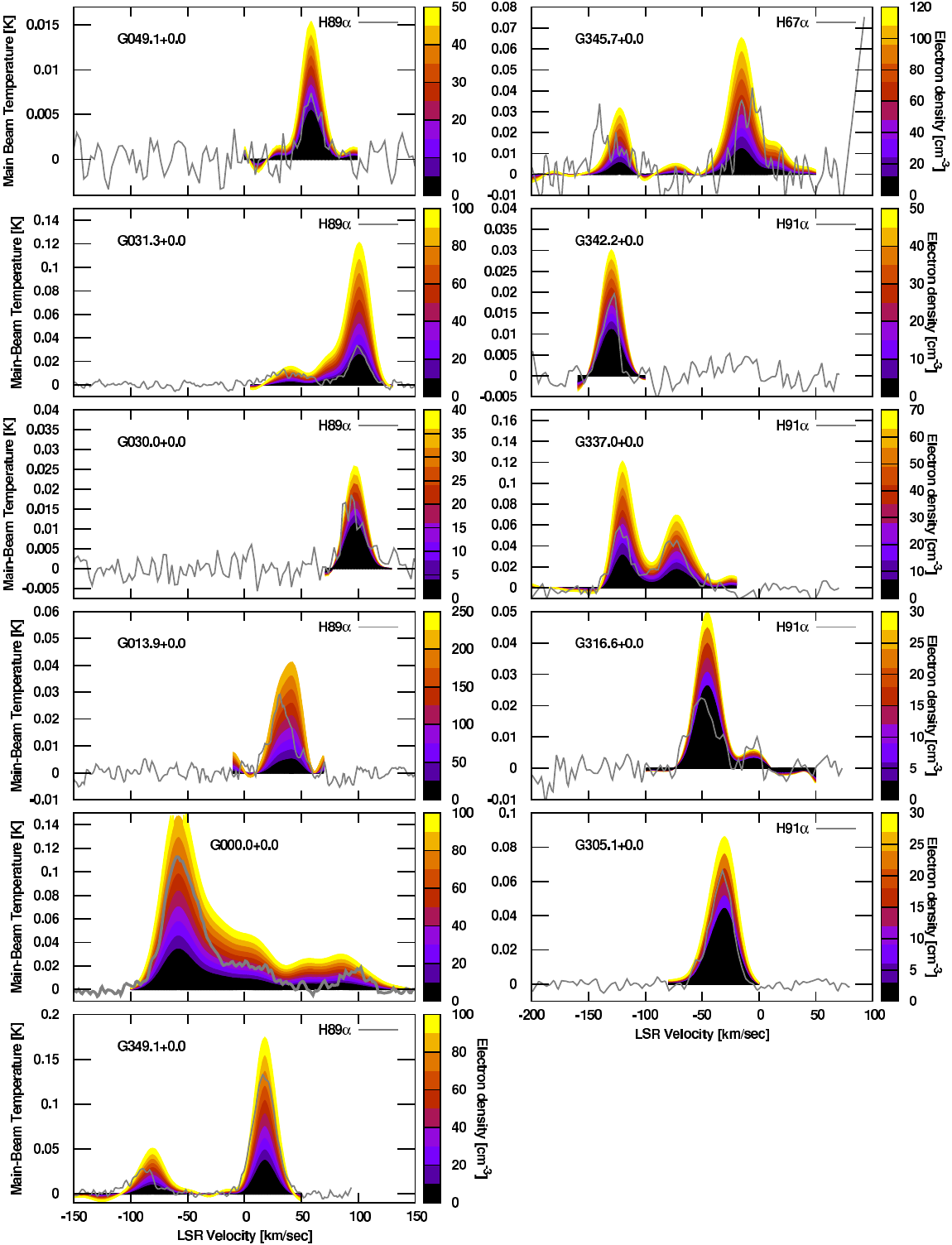}
\caption{Electron density solutions for our sample sight--lines. The
  grey lines show observed RRL emission. The color coded lines are the
  predicted RRL emission for the observed [N\,{\sc ii}] spectra and a
  given electron density (see Section~\ref{sec:results}), with the
  corresponding volume densities are shown in the color wedge on the
  right axis. Electron density solutions correspond to those where the
  observed RRL and its predicted emission coincide.
}\label{fig:density_results}
\end{figure*}

 \begin{figure}[t]
 \centering
 \includegraphics[width=0.45\textwidth,angle=0]{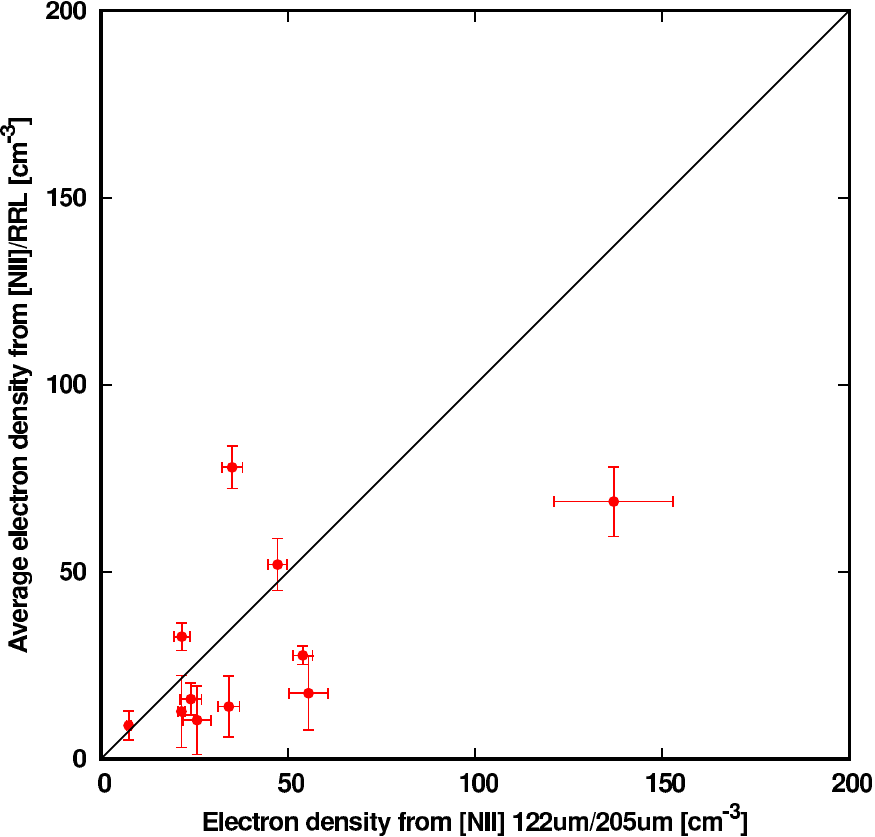}
 \caption{Electron density derived from the [N\,{\sc ii}]
   205$\mu$m/RRL ratio as a function of that derived from the [N\,{\sc
     ii}] 122$\mu$m/205$\mu$m ratio presented by
   \citet{Goldsmith2015}. The black straight line represent a
   $n_{e}^{\rm [NII]}=n_{e}^{\rm RRL}$
   relationship.}\label{fig:scatter_plot}
 \end{figure}

\subsection{Volume Densities}
\label{sec:results}

Using the method described in Section~\ref{sec:determ-electr-dens}, we
derived electron densities for 21 velocity components in the 11 LOSs
analyzed here. To determine the [N\,{\sc ii}]/RRL ratio for each
velocity components in the observed LOSs, we decomposed the [N\,{\sc
    ii}] and RRL emission into a set of Gaussian components
(Table\,\ref{tab:gauss-decomp-1}), which are later combined to
determine this line ratio.  We then use Equation~(\ref{eq:9}) to solve
for the electron density for each velocity component, with an assumed
electron temperature and nitrogen abundance relative to hydrogen, as
discussed above.  There are cases when the data are not easily
decomposed into Gaussian components due to blended components, in
particular for the RRL emission, as their line widths are
significantly larger than those for [N\,{\sc ii}]. In this case we
used the intensity integrated over a velocity range that encompasses
the velocity component for the derivation of the [N\,{\sc ii}]/RRL
ratio.

In Table~\ref{tab:densities}, we list the derived electron density,
LSR velocity, assumed electron temperature, nitrogen fractional
abundance, N$^+$ and H$^+$ column densities, continuum temperature,
and galactocentric distance for each velocity component analyzed here.
We find that the electron densities in our sample range between 8 and
170 cm$^{-3}$ with an average value of 41\,cm$^{-3}$.

%

In Figure~\ref{fig:density_results}, we illustrate our electron
density determination by showing the observed RRL emission together
with that predicted from the corresponding [N\,{\sc ii}] emission and
a set of electron densities (color--coded; see
Equation~\ref{eq:9}). Electron density solutions correspond to those
where the observed RRL and its predicted emission coincide. Note that
due to their different masses, the thermal line widths of H$^+$ are
expected to be significantly larger\footnote{For example, for
  $T_e$=8000\,K the thermal line width of a Hydrogen RRL line is 19
  km/sec while for the [N\,{\sc ii}] line it is 5 km\,s$^{-1}$. Note
  that pressure broadening is not expected to be significant in the
  density regime that we are sampling (\citealt{Brocklehurst1972}).  }
than those of N$^+$. To enable the comparison shown in
Figure~\ref{fig:density_results} we convolved the [N\,{\sc ii}] lines
with a Gaussian function with a FWHM that results in a [N\,{\sc ii}]
profile that matches the thermal line width for H$^+$.

\begin{deluxetable*}{lrrccccccccc} 
\tabletypesize{\footnotesize} \centering \tablecolumns{12} \small
\tablewidth{0pt} \tablecaption{Derived Electron Volume Densities  from the  [NII] 205$\mu$m/RRL  intensity ratio} \tablenum{3}
\tablehead{\colhead{LOS} & \colhead{$l$} & \colhead{$b$} & \colhead{V$_{\rm LSR}$} & \colhead{$n_{e}$} & \colhead{$N({\rm N^+})$} & \colhead{$N({\rm H^+})$} & \colhead{  $10^{4}$[{\rm N/H}]$^{1}$} & \colhead{$10^{4}$[{\rm N/H}]$^2$} & \colhead{$T_{\rm e}$} & \colhead{$R_{\rm gal}$} & \colhead{$T_{\rm C}$} \\ 
\colhead{}   & \colhead{[\degr]}   & \colhead{[\degr]}   & \colhead{[km\,s$^{-1}$]} & \colhead{[cm$^{-3}$]} &  \colhead{[$10^{16}$cm$^{-2}$]} & \colhead{[$10^{20}$cm$^{-2}$]} &  \colhead{} & \colhead{} &   \colhead{[K]} & \colhead{[kpc]} & \colhead{[K]}  \\}
\startdata 
G305.1+0.0 & 305.106 & 0.0 & $-$33.2 & 12.7$\pm$9.6 & 30.7 $\pm$ 16.6  &  37.9$\pm$32.3   & 0.77$\pm$0.06 & 0.6$\pm$0.1 & 7765$\pm$386 & 7.1  &  0.3 \\
G316.6+0.0 & 316.596 & 0.0 & $-$48.1 & 8.7$\pm$8.6 & 15.4 $\pm$ 3.6  &  11.3 $\pm$ 5.9   & 0.62$\pm$0.07 & 0.7$\pm$0.2 & 7439$\pm$371 & 6.4  &  0.2 \\
G316.6+0.0 & 316.596 & 0.0 & $-$6.4 & 14.1$\pm$12.0 & 7.8 $\pm$ 0.7  &  6.9 $\pm$ 2.5    & 0.62$\pm$0.07 & 0.5$\pm$0.1 & 8183$\pm$406 & 8.0  &  0.2 \\
G342.2+0.0 & 342.174 & 0.0 & $-$131.2 & 14.0$\pm$8.2 & 15.4 $\pm$ 3.6  &  11.3 $\pm$ 5.9  & 1.98$\pm$0.41 & 1.2$\pm$0.2 & 5504$\pm$311 & 2.3  &  0.1 \\
G337.0+0.0 & 336.957 & 0.0 & $-$121.5 & 28.0$\pm$10.3 & 2.4 $\pm$ 0.4  &  1.5 $\pm$ 0.9  & 0.74$\pm$0.09 & 1.1$\pm$0.2 & 5911$\pm$318 & 3.1  &  0.4 \\
G337.0+0.0 & 336.957 & 0.0 & $-$76.6 & 41.2$\pm$15.0 & 6.9 $\pm$ 0.4  &  20.0 $\pm$ 7.2  &0.74$\pm$0.09 & 0.9$\pm$0.2 & 6473$\pm$334 & 4.3  &  0.4 \\
G345.7+0.0 & 345.652 & 0.0 & $-$122.8 & 40.5$\pm$18.5 & 9.5 $\pm$ 1.0  &  28.2 $\pm$ 12.4 & --  & 1.3$\pm$0.3 & 5327$\pm$308 & 1.9  &  0.3 \\
G345.7+0.0 & 345.652 & 0.0 & $-$8.2 & 101.1$\pm$29.6 & 4.9 $\pm$ 0.6  &  3.1 $\pm$ 1.1    & --  & 0.6$\pm$0.1 & 7859$\pm$390 & 7.3  &  0.3 \\
G349.1+0.0 & 349.130 & 0.0 & 17.0 & 62.8$\pm$26.2 & 9.5 $\pm$ 1.0  &  28.2 $\pm$ 12.4 & 0.25$\pm$0.12  & 0.3$\pm$0.1 & 10600$\pm$540 & 13.2  &  0.3 \\
G349.1+0.0 & 349.130 & 0.0 & $-$91.1 & 42.0$\pm$13.0 & 4.9 $\pm$ 0.6  & 3.1 $\pm$ 1.1 & 1.16$\pm$0.62 & 1.2$\pm$0.2 & 5354$\pm$308 & 1.9  &  0.3 \\
G000.0+0.0 & 0.000 & 0.0 & $-$60.5 & 46.5$\pm$12.6 & 13.1 $\pm$ 1.0  &  8.0 $\pm$ 2.4  & 3.13$\pm$0.24 & 1.6$\pm$0.3 & 4446$\pm$301 & 0.0  &  7.0 \\
G000.0+0.0 & 0.000 & 0.0 & $-$37.1 & 14.6$\pm$7.2 & 28.2 $\pm$ 8.4  &  16.0 $\pm$ 9.1  & 3.13$\pm$0.24 & 1.6$\pm$0.3 & 4446$\pm$301 & 0.0  &  7.0 \\
G000.0+0.0 & 0.000 & 0.0 & 12.8 & 39.7$\pm$14.2 & 5.4 $\pm$ 1.1  &  3.3 $\pm$ 1.4 & 3.13$\pm$0.24 & 1.6$\pm$0.3 & 4446$\pm$301 & 0.0  &  7.0 \\
G000.0+0.0 & 0.000 & 0.0 & 95.0 & 11.3$\pm$6.5 & 10.6 $\pm$ 4.1  &  6.0 $\pm$ 4.0 & 3.13$\pm$0.24 & 1.6$\pm$0.3 & 4446$\pm$301 & 0.0  &  7.0 \\
G013.9+0.0 & 13.913 & 0.0 & 45.5 & 28.6$\pm$12.3 & 1.3 $\pm$ 0.3  &  1.1 $\pm$ 0.7    & 1.80$\pm$0.51     & 0.9$\pm$0.2 & 6511$\pm$336 & 4.4  &  0.4 \\
G013.9+0.0 & 13.913 & 0.0 & 30.3 & 166.3$\pm$44.2 & 0.7 $\pm$ 0.1  &  0.8 $\pm$ 0.2   & 1.80$\pm$0.51    & 0.8$\pm$0.2 & 6944$\pm$351 & 5.3  &  0.4 \\
G030.0+0.0 & 30.000 & 0.0 & 95.4 & 17.6$\pm$9.9 & 6.3 $\pm$ 2.2  &  5.4 $\pm$ 3.6     & 1.91$\pm$0.35      & 0.9$\pm$0.2 & 6502$\pm$336 & 4.4  &  0.6 \\
G031.3+0.0 & 31.277 & 0.0 & 100.4 & 7.9$\pm$7.9 & 31.9 $\pm$ 29.8  &  26.6 $\pm$ 31.3 & 1.04$\pm$0.13 & 0.9$\pm$0.2 & 6488$\pm$336 & 4.4  &  0.4 \\
G031.3+0.0 & 31.277 & 0.0 & 38.0 & 108.8$\pm$46.6 & 1.6 $\pm$ 0.1  &  1.9 $\pm$ 0.7   & 1.04$\pm$0.13 & 0.7$\pm$0.2 & 7407$\pm$369 & 6.3  &  0.4 \\
G049.1+0.0 & 49.149 & 0.0 & 59.7 & 10.4$\pm$9.2 & 3.8 $\pm$ 2.6  &  4.2 $\pm$ 4.1     & 1.03$\pm$0.25      & 0.7$\pm$0.2 & 7351$\pm$367 & 6.2  &  0.2 \\
\enddata
\tablenotetext{1}{Nitrogen abundance relative to hydrogen derived by matching electron densities derived in our work with those derived from the [N\,{\sc ii}] 122$\mu$m/205$\mu$m ratio. }
\tablenotetext{2}{Nitrogen abundance relative to hydrogen derived from the fit presented  by \citet{Esteban2018}. }
\label{tab:densities}
\end{deluxetable*}

\begin{deluxetable}{lccccccc} 
\tabletypesize{\footnotesize} \centering \tablecolumns{3} \small
\tablewidth{0pt} \tablecaption{LOS Averaged Volume Densities} \tablenum{4}
\tablehead{\colhead{LOS} &  \colhead{$n_{e}\left (\frac{{\rm [NII]} 205\mu m}{\rm RRL} \right )$} & \colhead{$n_{e}\left(\frac{[\rm NII] 122\mu m}{[\rm NII] 205\mu m} \right )$}    \\ 
\colhead{}  & \colhead{[cm$^{-3}$]} &   \colhead{[cm$^{-3}$]}  \\}
\startdata
%
G305.1+0.0 &  12.7$\pm$9.6 & 21.5$\pm$0.9    \\ 
G316.6+0.0 &  8.9$\pm$3.8 & 7.4$\pm$0.6      \\ 
G337.0+0.0 &  32.7$\pm$3.7 & 21.6$\pm$2.2    \\ 
G342.2+0.0 &  14.0$\pm$8.2 & 34.2$\pm$2.9    \\ 
G345.7+0.0 &  77.9$\pm$5.7 & 35.1$\pm$2.8    \\ 
G349.1+0.0 &  52.0$\pm$7.0 & 47.2$\pm$2.6    \\ 
G000.0+0.0 &  27.7$\pm$2.5 & 54.0$\pm$2.6    \\ 
G013.9+0.0 &  68.8$\pm$9.3 & 137.1$\pm$15.9  \\ 
G030.0+0.0 &  17.6$\pm$9.9 & 55.5$\pm$5.2    \\ 
G031.3+0.0 &  16.0$\pm$4.3 & 24.0$\pm$2.9    \\ 
G049.1+0.0 &  10.4$\pm$9.2 & 25.7$\pm$3.7    \\ 
\enddata
\tablenotetext{}{}
\label{tab:pacs}
\end{deluxetable}

\subsection{Comparison with densities derived using 122$\mu$m/250$\mu$m ratio}
\label{sec:comp-with-dens}

\citet{Goldsmith2015} used low spectral resolution {\it Herschel}/PACS
observations of the [N\,{\sc ii}] 122$\mu$m and 205$\mu$m lines to derive a
characteristic electron density along the LOSs across the Galactic
plane, including the 11 LOSs studied here. The ratio of the [N\,{\sc
    ii}] 122$\mu$m to 205$\mu$m ratio provides an accurate
determination of the electron density that does not depend on nitrogen
abundance, and has a weak dependence on electron temperature. A
comparison with our velocity--resolved determination of the electron
density provides an opportunity to understand how this characteristic
density determined by {\it Herschel}/PACS relates to the actual volume
density distribution along the line--of--sight and also allows us to
test our approach to determine the electron density.

For the low density regime, the electron density derived from the
integrated [N\,{\sc ii}] 122$\mu$m to 205$\mu$m ratio is the nitrogen
column density weighted average of all components along the
line--of--sight. However, for the density regime found in our LOSs the
electron density has a less straightforward interpretation. 

The [N\,{\sc ii}] 122$\mu$m to 205$\mu$m ratio for a LOS with $n$
velocity components with integrated intensities $I^i_{\rm 122\mu m}$
and $I^j_{\rm 205\mu m}$ is given by,

\begin{equation}\label{eq:13}
\frac{I_{\rm 122\mu m}}{I_{\rm 205 \mu m}}=\frac{\sum^n_i I^i_{\rm 122\mu m}}{\sum^n_j I^j_{\rm 205 \mu m}}.
\end{equation}

Substituting Equation~(\ref{eq:1}) in the numerator of
Equation~(\ref{eq:13}) and considering that the upper level column
density, $N_u$, is equal to the ionized nitrogen column density times
the fractional abundance of the upper level, $N(\rm N^+)f_2 (n_{e})$, results
in,

\begin{equation}\label{eq:19}
\frac{I_{\rm 122\mu m}}{I_{\rm 205 \mu m}}=\frac{A_{21} h \nu_{\rm 122 \mu m}} {4\pi}\frac{\sum^n_i N^i({\rm N^+}) f^{i}_2(n^i_{\rm e}) }{\sum^n_j I^j_{\rm 205 \mu m}}.
\end{equation}

Also using Equation~(\ref{eq:1}), we can express the ionized nitrogen
column density as a function of the [N\,{\sc ii}] 205$\mu$m line
intensity (for which we have information about its velocity
distribution) as, 
\begin{equation}\label{eq:14}
N({\rm N^+})= \frac{4\pi I_{205 \mu m}}{A_{10} h \nu_{205 \mu m} f_1(n_e)}.
\end{equation}

Substituting this expression in Equation~(\ref{eq:19}) allows us to write
the [N\,{\sc ii}] 122$\mu$m/205$\mu$m ratio as a function of the
[N\,{\sc ii}] 205\,$\mu$m intensity and the $f_1(n_e)$ and $f_2(n_e)$
upper level fractions as,
\begin{equation}\label{eq:15}
\frac{I_{\rm 122\mu m}}{I_{\rm 205 \mu m}}=\frac{A_{21} \nu_{\rm 122 \mu m}}{A_{10} \nu_{\rm 205\mu m}}
\frac{\sum^n_i I^i_{\rm 205 \mu m} \frac{ f^{i}_2(n^i_{\rm e})}{f^{i}_1(n^i_{\rm e})}}{\sum^n_j I^j_{\rm 205 \mu m}},
\end{equation}
and thus can be evaluated given the electron densities derived from
the [N\,{\sc ii}]/RRL ratio and the observed integrated intensities of
the [N\,{\sc ii}] 205$\mu$m line in each velocity component along a
given LOS. This estimated ratio can be used to evaluate the
characteristic electron density using Equation 21 in
\citet{Goldsmith2015}. Note that the electron densities published by
\citet{Goldsmith2015} are those for the [N\,{\sc ii}] 122$\mu$m to
205$\mu$m intensities averaged over the 47\arcsec\ footprint of the
{\it Herschel}/PACS instrument. In our calculations we use instead, to
avoid uncertainties due to beam dilution, the {\it Herschel}/HIFI
spectra at a resolution of 16\arcsec\ and the [N\,{\sc ii}] 122$\mu$m
intensities averaged from the {\it Herschel}/PACS footprint with a
Gaussian weighted function with a FWHM of 16\arcsec. These different
approaches result in slightly different electron densities.

In Table~\ref{tab:pacs}, we compare the characteristic electron
density derived from Equation~(\ref{eq:15}) together with that
resulting from the observed [N\,{\sc ii}] 122$\mu$m to 205$\mu$m ratio
by {\it Herschel}/PACS.  In Figure~\ref{fig:scatter_plot}, we compare
the LOS-averaged electron density with those derived using the
122$\mu$m/205$\mu$m lines by \citet{Goldsmith2015}. While there is
general agreement between these two methods, the scatter can arise
from the assumed nitrogen fractional abundance in our calculations.

\subsection{Nitrogen Abundances relative to Hydrogen}
\label{sec:nitr-abund-relat}

\begin{figure*}[t]
\centering
\includegraphics[width=0.7\textwidth,angle=0]{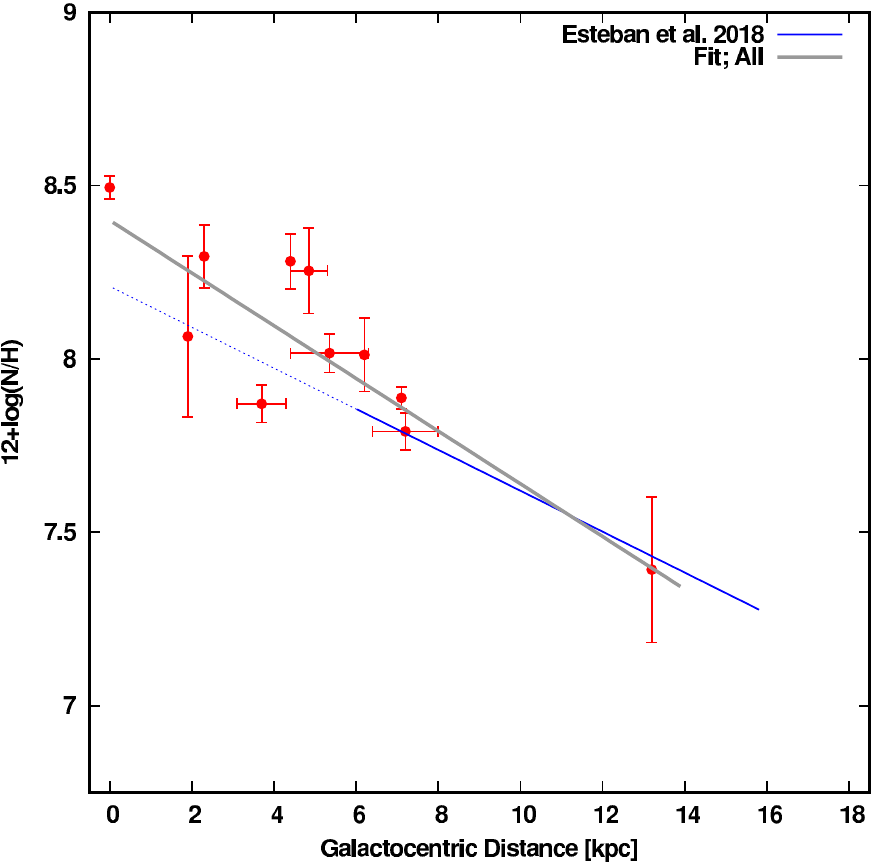}
\caption{Nitrogen abundance relative to hydrogen as a function of
  Galactocentric distance. The blue straight line represents the fit
  to optical data from \citet{Esteban2018} and an extrapolation of
  this relationship to $R_{\rm gal}$=0\,kpc is shown as a blue dashed
  line. The fit to our data is shown as a grey
  line.}\label{fig:Xn_calculation}
\end{figure*}

The distribution of element abundances in the disk of galaxies is a
fundamental observational constraint for models of the formation and
evolution of the Milky Way. Nitrogen is formed in ``primary" and
"secondary" processes in massive and intermediate mass stars
\citep{Johnson2019}, and thus its abundance distribution is related to
the star formation history as a function of Galactocentric distance in
the Milky Way.  Observations of optical lines of [N\,{\sc ii}] in
nearby H\,{\sc ii} regions, where dust extinction obscure these lines
moderately, have been used to infer that the abundance of Nitrogen
increases with Galactocentric distance from the outer galaxy up to
$R_{\rm gal}= 6$\,kpc \citep{Esteban2018}. Due to increased dust
extinction, optical studies are however unable to probe the inner
Galaxy, where stellar processing was enhanced in the past.  The
far--infrared lines of [N\,{\sc ii}] have the advantage to be
unobscured, and thus allow us to determine element abundances over
larger distances, including the inner galaxy, where most of the star
formation takes place \citep{Simpson1995,Simpson2004}.
In the following, we use our observations of RRLs together with those
of the [N\,{\sc ii}] 205$\mu$m and 122$\mu$m lines to derive nitrogen
abundances in our sample allowing us to sample the inner Galaxy.

The [N\,{\sc ii}] 122$\mu$m/205$\mu$m ratio provides an accurate
determination of the electron density that is independent of nitrogen
abundance and has a weak dependence of electron temperature.  By
matching electron temperatures derived from the [N\,{\sc ii}]
205$\mu$m/RRL ratio to those derived from the [N\,{\sc ii}]
122$\mu$m/205$\mu$m ratio, we can therefore determine the abundance of
ionized nitrogen relative to ionized hydrogen in our sample.

The ionized nitrogen abundance, $X_{\rm N^+}$, can be written in terms
of the ionized nitrogen column density, $N({\rm N^+})$, and the
emission measure, $EM$, derived from the RRL observations
(Equations~\ref{eq:4} and \ref{eq:5}) as,
\begin{equation}\label{eq:10}
X_{\rm N^+} =\frac{N({\rm N^+})}{N({\rm H^+})}=\frac{n_e N({\rm N^+})}{EM},
\end{equation}
which in turn can be written in terms of the [N\,{\sc ii}] 205$\mu$m
line intensity using Equation~(\ref{eq:14}) as,
\begin{equation}\label{eq:16}
X_{\rm N^+}=\frac{4\pi n_e I_{205 \mu m}}{A_{10} h \nu_{205 \mu m} f_1(n_e)EMG^{-1}_{\rm NLTE}(n_e,T_e)},
\end{equation}

Using this Equation, we determined the ionized nitrogen abundance in
LOSs with a single velocity component (three in our sample) by using
the observed $I_{205 \mu m}$ and $EM$, and assuming an electron
density that results from the [N\,{\sc ii}] 122$\mu$m/205$\mu$m ratio
observed with {\it Herschel}/PACS.

For spectra with two velocity components we find that there is
typically a large number of combinations of the electron density of
each component that results in the same observed [N\,{\sc ii}]
122$\mu$m/205$\mu$m ratio (see Equation~\ref{eq:15}), and thus the
$X_{\rm N^+}$ of each individual component cannot be uniquely
determined. However, the velocity of the components in most of these
LOSs (5 out of 7) are similar to each other and have distances to the
Galactic center that are within 2\,kpc from each other.   Assuming
  that $X_{\rm N^+}$ varies only over large scales, we can therefore
  use a constant value of $X_{\rm N^+}$ for sources within $<$2\,kpc
  scales.  In this case, the electron density, $n^i_e$, of any
component is related to that, $n^j_e$, of any other component by,
\begin{equation}\label{eq:17}
  \frac{n^i_e I^i_{205 \mu m}}{f_1(n^i_e)EM^i G^{-1}_{\rm NLTE}(n^i_e,T^i_e)}=\frac{n^j_e I^j_{205 \mu m}}{f_1(n^j_e)EM^j G^{-1}_{\rm NLTE}(n^j_e,T^j_e)},  
\end{equation}
and this expression can be used in combination with
Equation~(\ref{eq:15}) to solve for the electron densities for all
velocity components. This information can be then used to determine
$X_{\rm N^+}$ for the LOS using Equation~(\ref{eq:16}).

There are two LOSs in our sample, G345.7+0.0 and G349.1+0.0, for which
we cannot assume a constant, $X_{\rm N^+}$, as their velocity
components are well separated, corresponding to Galactocentric
distances that differ by more than 5\,kpc. In the case of G349.1+0.0
the electron density solutions obtained using the [N\,{\sc ii}]
205$\mu$m/RRL also satisfy Equation~(\ref{eq:15}) within their
uncertainties and we therefore assume a solution for the electron
densities that is the closest electron density combination that can
satisfy both the observed [N\,{\sc ii}] 205$\mu$m/RRL and [N\,{\sc
  ii}] 122$\mu$m/205$\mu$m ratios.  This similarity between solutions
is not the case for G345.7+0.0 and therefore we excluded this LOSs
from our estimate of $X_{\rm N^+}$.

In Figure~\ref{fig:Xn_calculation} we present the derived ionized
nitrogen fractional abundances as a function of Galactocentric
distance. We also include a straight solid line corresponding to the
fit by \citet{Esteban2018}, which is determined for a range of 6\,kpc
to 16\,kpc in Galactocentric distance. We also include the
extrapolation of this relationship to $R_{\rm gal}$=0\,kpc as a dashed
line. Our determination of $X_{\rm N^+}$ corresponds mostly to the
range $R_{\rm gal}<6$\,kpc which is not accessible for optical studies
such as that from \citet{Esteban2018}.
 We fitted a straight line to all data in
  Figure~\ref{fig:Xn_calculation} using the orthogonal bi-variate
  error and intrinsic scatter method (BES\footnote{Available at
    \url{http://www.astro.wisc.edu/~mab/archive/stats/stats.html}}
  ; \citealt{Akritas1996}), including a bootstrap resampling error
  analysis, resulting in,
\begin{equation}\label{eq:18}
12+\log({\rm N}^+/{\rm H}^{+})=8.40(\pm0.21)-0.076(\pm0.034)R_{\rm gal}.
\end{equation}
 We checked whether this fit was influenced by the single data
  point at 13.2\,kpc, by re-fitting the data with all data but
  excluding this data point. We found no significant difference
  between the resulting fit and that found for the entire data set.
%
%
%
%
%
We find that our derived ionized nitrogen abundances have a slope that
is consistent with that found by \citet{Esteban2018} in the outer
Galaxy. A flattening in the abundance distribution of different
species in the inner galaxy have been inferred by observations of
Cepheids and red giants \citep{Hayden2014,Martin2015,Andrievsky2016},
suggesting different star formation histories between the inner and
outer Galaxy, with the star formation rate in the inner galaxy
possibly being reduced by dynamical effects such as the Galactic
bar. While the distribution of nitrogen abundances in our data is not
consistent with the suggested flattening in the inner Galaxy, it
cannot be discarded with our data due to the uncertainties in the fit,
which in turn results from the small number of samples. As mentioned
above, we are currently conducting a survey of all GOT\,C+ LOSs
observed in [N\,{\sc ii}] by \citet{Goldsmith2015}, with the GBT and
DSS--43 telescopes, which will allow us to increase significantly our
sample in the inner galaxy, resulting in an accurate determination of
the distribution of nitrogen abundances in the inner Galaxy.

In H\,{\sc ii} regions, nitrogen is efficiently singly ionized by
charge exchanges with protons \citep{Langer2015b}, collisional
ionization with electrons, and an extreme ultraviolet (EUV)
field. Higher ionization states of nitrogen are possible in the
presence of a large EUV fields in the close vicinity of massive star
clusters.  Higher ionization states of nitrogen need to be accounted
for a determination of the total nitrogen abundance, $X_{\rm N}$, from
our derived ionized nitrogen abundance. The abundance of higher
ionization states of nitrogen can be estimated with observations of
[N\,{\sc iii}] and [O\,{\sc iii}] lines (N$^+$ and O$^+$ have similar
ionization potentials; 29.6 and 35.1 eV, respectively). But these
observations are not always available.  \citet{Esteban2018} observed
optical collisonally excited lines of N in low--ionization H\,{\sc ii}
regions, where the abundance of multiple ionized nitrogen is
negligible, and thus $X_{\rm N} \simeq X_{\rm N^+}$. As mentioned in
Section~\ref{sec:beam-dilut-effects}, except for G345.7+0.0 and
G349.1+0.0, our LOSs are not directly associated with the brightest
parts of known H\,{\sc ii} regions, and represent the extended
envelopes of these H\,{\sc ii} regions. Additionally, the derived
electron densities are lower than those typically found in compact
H\,{\sc ii} regions. We can therefore assume that the ionized gas
probed in our sample is also low--ionization, and most of the
gas--phase nitrogen is singly ionized. Note that even if an accounting
for multiple ionization states of nitrogen is needed, we do not expect
a significant variation of this contribution to the total nitrogen
abundance to vary with Galactocentric distance, and thus the slope the
relationship between our derived nitrogen abundances and
Galactocentric distance should not vary significantly.

\subsection{The nature of the dense Warm ionized medium}
\label{sec:nature-dense-warm}

The derived densities in our sample confirm those derived by
\citet{Goldsmith2015} over a larger sample in the Galactic plane, but
at lower spectral resolution. The observed [N\,{\sc ii}] and RRL
emission arise from a plasma that is denser than the diffuse warm
ionized medium ($\lesssim$0.1\,cm$^{-3}$;
\citealt{Cordes2002,Haffner2009}) but has lower densities compared to
those typical of (ultra-) compact H\,{\sc ii}
regions\citep[$>5\times10^3 {\rm cm}^{-3}$;][]{Kurtz2005}.  The
spectrally resolved observations allowed us to study the electron
density of these media along the line of sight, showing that the RRL
and [N\,{\sc ii}] emission arise from discrete velocity components
rather than from the extended diffuse warm ionized medium, which would
show emission over a much wider range of velocities.  By comparing
with the WISE catalog of known H\,{\sc ii} regions
(Table~\ref{tab:xmatch3}, Figure~\ref{fig:environment}), we find that
most of our LOSs are located in the vicinity of H\,{\sc ii} regions
but do not overlap with their brightest parts, and thus are unlikely
to be associated with compact H\,{\sc ii} regions.  This dense plasma
appears to be widely distributed in the Galactic plane, as suggested
by similar densities detected in a larger number of LOSs by
\citet{Goldsmith2015} and at the edge of the central molecular zone by
\citet{Langer2015b}, and thus can represent a significant fraction of
the ionized gas in our Galaxy. Further investigations of this
widespread dense ionized gas component, by large scale mapping of
[N\,{\sc ii}] and RRL emission, are important for fully characterizing
this component and to assist the interpretation of observations of
ionized gas in extra-galactic sources.

\section{Conclusions}
\label{sec:conclusions}

We presented a method to derive the electron density of ionized gas
using the ratio of the [N\,{\sc ii}] 205$\mu$m line to a radio
recombination line. We use this method to derive electron densities of
21 velocity components in 11 LOSs observed in spectrally resolved
[N\,{\sc ii}] 205$\mu$m with the {\it Herschel}/HIFI and SOFIA/GREAT
instruments and in radio recombination lines with the Green Bank
Telescope and NASA Deep Space Network DSS-43 telescope. We summarize
our results as follows:

   \begin{itemize}

   \item We found typical electron densities between 6--170\,cm$^{-3}$
     with an average value of 41\,cm$^{-3}$, which are consistent to
     those derived at low spectral resolution using the [N\,{\sc ii}]
     122$\mu$m/205$\mu$m ratio with {\it Herschel}/PACS, and are
     significantly larger than those characteristic of the diffuse
     Warm Ionized Medium but lower than those expected for (ultra-)
     compact H\,{\sc ii} regions.

   \item By matching the electron densities derived from the [N\,{\sc
       ii}] 205$\mu$m/RRL ratio and the [N\,{\sc ii}]
     122$\mu$m/205$\mu$m ratio, we derive the fractional abundance N/H
     for most of our velocity components.

   \item We studied the dependence of the N/H ratio with
     Galactocentric distance in the inner 8\,kpc of the Milky Way,
     which is a range that is not accessible (in particular for
     $R_{\rm gal}<6$\,kpc) to optical studies due to dust
     extinction. We find that the distribution of nitrogen abundances
     in the inner galaxy derived from our data has a slope that is
     consistent with that found in the outer Galaxy in optical
     studies. This result is inconsistent with suggestions of a
     flatter distribution of the nitrogen abundance in the inner
     galaxy.  This trend, however, will need to be confirmed with a
     larger [N\,{\sc ii}] and RRL data set.

   \end{itemize}
   The method presented here to derive electron densities can be
   expanded to a much larger data base using the {\it Herschel}/PACS
   [N\,{\sc ii}] 122$\mu$m and 205$\mu$m survey by
   \citet{Goldsmith2015} consisting in about 100 LOSs and our complete
   DSS--43/GBT survey of the Galactic plane. These observations will
   allow us to obtain a more complete sample of electron densities in
   the Galactic plane and to derive the N/H gradient in our Galaxy.

\begin{acknowledgements}

  This research was conducted at the Jet Propulsion Laboratory,
  California Institute of Technology under contract with the National
  Aeronautics and Space Administration.  This project made use of the
  Smithsonian Astrophysical Observatory $4 \times
  32\mathrm{k}$-channel spectrometer (SAO32k) and the \emph{TAMS
    observatoryCtrl} observing system, which were developed by L.
  Greenhill (Center for Astrophysics), I. Zaw (New York University Abu
  Dhabi), D. Price, and D. Shaff, with funding from SAO and the NYUAD
  Research Enhancement Fund and in-kind support from the Xilinx
  University Program. We thank West Virginia University for its
  financial support of GBT operations, which enabled the observations
  for this project. The National Radio Astronomy Observatory is a
  facility of the National Science Foundation operated under
  cooperative agreement by Associated Universities, Inc.  LDA and ML
  are supported by NSF grant AST1516021 to LDA.
\copyright\ 2019 California
  Institute of Technology. U.S. Government sponsorship acknowledged.

\end{acknowledgements}
\software{TMBIDL \citep{Bania2014}, GILDAS/CLASS \citep{Pety2005}}
\facilities{GBT, DSN/DSS--43, SOFIA, Herschel}

\appendix
\section{Beam Dilution Correction}
\label{sec:beam-dilut-corr}

Our analysis is carried out using data at different angular
resolutions (16\arcsec\ for [N\,{\sc ii}] and 84\arcsec\ or
115\arcsec\ for the RRL data), and therefore our derived electron
densities can be affected by beam dilution effects.  As a first step
to address the effects of beam dilution, we checked whether any of our
LOSs are associated with known H\,{\sc ii} regions in the WISE catalog
(Table~\ref{tab:xmatch3} and Figure~\ref{fig:environment}). We found
that most of our LOSs are not directly associated with compact H\,{\sc
  ii} regions but are located in their vicinity. There are two LOSs in
our sample (G345.7+0.0 and G349.1+0.0) in which the RRL beam partially
overlap with H\,{\sc ii} regions in the WISE catalog. In the following
we investigate how the results of this paper are affected by beam
dilution effects.  We studied beam dilution effects in the [N\,{\sc
  ii}] data using {\it Herschel}/PACS data, and in the RRL data using
radio continuum data.  The dilution factors derived from these data
sets, and their impact in the derived electron density in each
velocity component, are summarized in Table~\ref{tab:dilution}.

While the {\it Herschel}/HIFI [N\,{\sc ii}] data have an angular
resolution of 15.7\arcsec, we also have observations of the same line
done with the PACS instrument which has a $5\times5$ pixel grid, with
a pixel separation of 9.4\arcsec, corresponding to a footprint of
47\arcsec\ in the sky. Note however, that the PACS data is velocity
unresolved, and thus provides information about the integrated
intensity of the [N\,{\sc ii}] over 47\arcsec\ scales. To study the
beam dilution of the [N\,{\sc ii}] line over 47\arcsec\ scales we used
the PACS [N\,{\sc ii}] footprint to calculate the intensity at
15.7\arcsec\ and 47\arcsec\ by averaging the data weighted by a two
dimensional Gaussian function with FWHM corresponding to 15.7\arcsec\
and 47\arcsec, respectively.  We find a typical variation of the
[N\,{\sc ii}] intensity of about 15\% on angular scales extending from
15.7\arcsec\ to 47\arcsec. We used the derived beam dilution factors
to correct our data using the method described in Appendix A.2 in
\citet{Pineda2017}, with the exception of G000.0+0.0 in which we have
data available at 16\arcsec\ resolution, as discussed below.

 We also studied beam dilution effects over a larger range of
  angular resolutions, by using radio continuum observations from the
  THOR+VGPS ($\lambda=$20\,cm; angular resolution 25\arcsec; coverage
  $l=14.5$\degr\ to 67.4\degr; \citealt{Beuther2016}), MAGPIS (20\,cm;
  6\arcsec ; $l=5$\degr\ to 48.5\degr; \citealt{Helfand2006}), and
  MGPS--2 (35.6\,cm; 45\arcsec--57.5\arcsec; $l=245$\degr\ to
  365\degr; \citealt{Murphy2007}) surveys. The THOR+VGPS and MAGPIS
  maps have been corrected for missing short spacings, and the MGPS--2
  data is sensitive to angular scales between 45\arcsec and
  1200\arcsec--1800\arcsec, thus encompassing the angular scales
  sampled by the {\it Herschel}/PACS, GBT, and DSS--43
  observations. We used the THOR+VGPS data of our sample with
  $l>14.5\degr$, the MAGPIS data for G013.9+0.0, and the MGPS--2 data
  for sources with $l<365\degr$. We assume that, given that RRL
  emission is detected in our sample, we assume that radio continuum
  emission is dominated by thermal Bremsstrahlung (free--free)
  emission from the ionized gas rather than from Synchrotron
  radiation, and thus its morphology corresponds to that of the
  ionized gas.  We convolved these map to 47\arcsec\ (except for the
  MGPS2 data for which we use its original resolution), 84\arcsec, and
  115.2\arcsec\ to match that of the {\it Herschel}/PACS, GBT, and
  DSS--43 data, respectively.  We find that intensities vary typically
  by 6\% when smoothing from 47\arcsec\ to 84\arcsec\, and by 14\%
  when smoothing from 47\arcsec\ to 115\arcsec. These variations
  results in typical variations in the derived electron densities of
  about 38\%. Note that two LOSs, G337.0+0.0 and G342.2+0.0, are not
  detected in radio continuum emission in the MGPS2 survey. They both
  are adjacent to extended H\,{\sc ii} regions
  (Figure~\ref{fig:environment}) that do not overlap significantly
  with the {\it Herschel}/PACS footprint and DSS--43 beam. We
  therefore assume that there is no significant spatial variations
  between 47\arcsec and 115\arcsec\ in these LOSs and we thus use a
  dilution factor of unity for these sources.  By applying these
  dilution factors to our data we are estimating electron densities on
  scales of 47\arcsec.

\begin{figure}[t]
\centering
\includegraphics[width=\textwidth,angle=0]{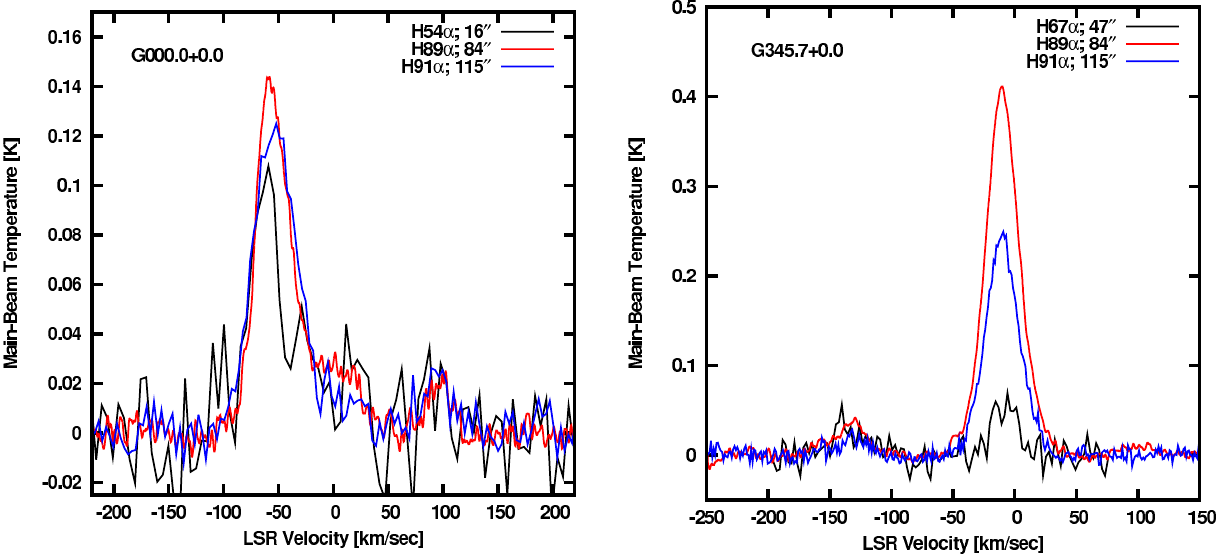}
\caption{({\it left panel}) The G000.0+0.0 line--of--sight observed in
  the H54$\alpha$, H89$\alpha$, and H91$\alpha$, at 16\arcsec,
  84\arcsec, and 115\arcsec angular resolution, respectively. ({\it
    right panel}) The G345.70+0.0 line--of--sight observed in the
  H67$\alpha$, H89$\alpha$, and H91$\alpha$, at 47\arcsec, 84\arcsec,
  and 115\arcsec angular resolution, respectively.  All RRL
  intensities shown here were are scaled to correspond to that of the
  H89$\alpha$ line.}\label{fig:results_gc}
\end{figure}

  We made a better estimate of the beam dilution effects in two LOSs
  by comparing their RRL observations taken at different angular
  resolutions. In the left panel of Figure~\ref{fig:results_gc}, we
  show the G000.0+0.0 LOS observed in the H54$\alpha$ and H89$\alpha$
  lines with the GBT at 16\arcsec\ and 84\arcsec, respectively, and in
  the H91$\alpha$ line observed with the DSS-43 telescope at
  115\arcsec.  All RRLs observations shown in
  Figure~\ref{fig:results_gc} in the plot were scaled to correspond to
  the intensity of the H89$\alpha$ line.  As discussed in
  Section~\ref{sec:hydr-recomb-lines}, when scaling RRLs with large
  differences between their principal quantum numbers, deviations from
  LTE, which depend on electron density and temperature
  (Equation~\ref{eq:7}), need to be taken into account.  We therefore
  considered non--LTE effects when scaling the H54$\alpha$ line
  intensities to correspond to that of the H89$\alpha$ line, by
  assuming the electron density and temperatures shown in Table 3 for
  G000.0+0.0.  Note that the contribution from non--LTE effects to the
  scaling factor between H54$\alpha$ and H89$\alpha$ varies weakly
  with electron density in the uncertainty range in this quantity that
  result from beam dilution effects and other error sources
  (Section~\ref{sec:uncertainties}).  While the velocity distribution
  is similar at these different angular resolution, we find that the
  84\arcsec\ H89$\alpha$ data is a factor of 1.5 times brighter than
  that at 16\arcsec. Given that the H89$\alpha$ observations have the
  highest signal--to--noise ratio, we will use this spectrum for our
  analysis but with the factor 1/1.5 applied to account for the beam
  dilution effects in our data.  In the right panel of
  Figure~\ref{fig:results_gc}, we show the H67$\alpha$, H89$\alpha$,
  and H91$\alpha$ lines, at 47\arcsec, 84\arcsec, and 115\arcsec,
  respectively, in the G345.7+0.0 LOSs.  As above, we accounted for
  non--LTE effects when scaling the H67$\alpha$ line intensity to that
  of the H89$\alpha$ line.  This LOS is associated with a bright
  H\,{\sc ii} region found in the WISE catalog (ID: G345.651+00.015;
  see also \citealt{Caswell1987}). This source is 55.5\arcsec\ away
  from our pointing direction, and thus it overlaps with both the GBT
  and DSS-43 beams, but not with the {\it Herschel}/HIFI and PACS
  observations (see Figure~\ref{fig:environment}). The data at
  47\arcsec\ is a factor of $\sim$0.2 and $\sim$0.13 fainter that
  those observed by the DSS--43 and GBT, respectively, and is thus
  relatively unaffected by the G345.651+00.015 source. Given that we
  have [N\,{\sc ii}] observations at 47\arcsec\ resolution, we use the
  47\arcsec\ H67$\alpha$ data in our analysis. Note that
  G345.651+00.015 shows only emission for the component at
  $-$8\,km\,s$^{-1}$ \citep{Caswell1987}, and thus beam dilution
  affects this velocity component only. The emission for the velocity
  component at $-$122.8\,km\,s$^{-1}$ appears to be consistent for all
  observed LOSs, and thus for our analysis we will use the observation
  at H91$\alpha$ which has a higher signal--to-noise ratio.

\begin{figure*}[t]
\centering
\includegraphics[width=\textwidth,angle=0]{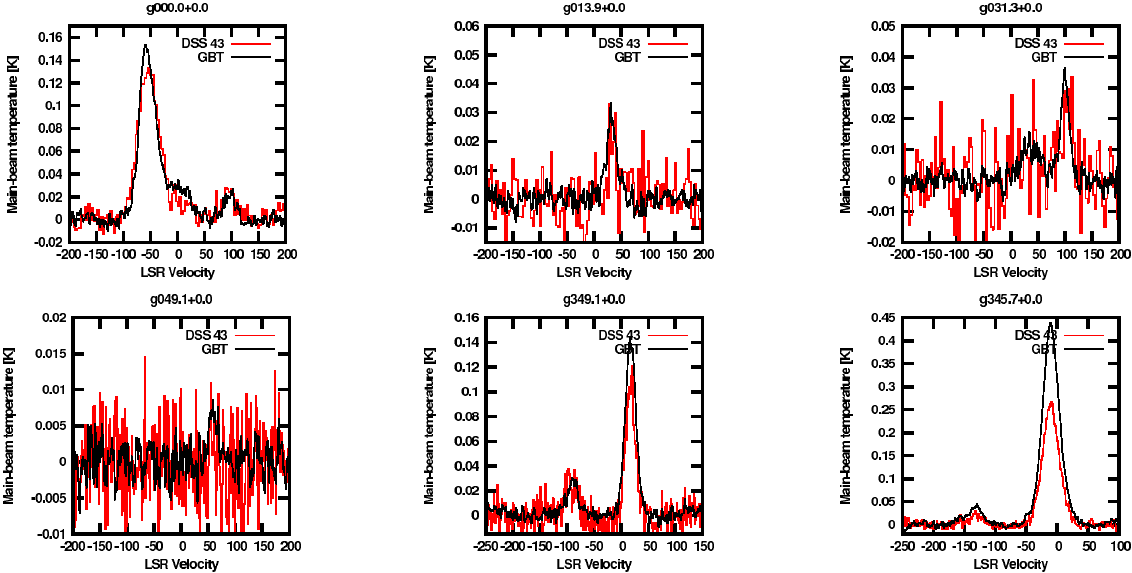}
\caption{Comparison between hydrogen recombination line observations
  from the GBT and the DSS--43 telescopes. We find good agreement
  between these two lines, despite the different angular resolution
  of the observations (115\arcsec\ vs 84\arcsec).
}\label{fig:comparison_data}
\end{figure*}

  In Figure~\ref{fig:comparison_data} we show a comparison in a larger
  number of LOSs between the GBT and DSS--43 observations. Apart from
  the LOSs discussed above, we find general agreement between
  observations taken with these two telescopes. All RRLs in the plot
  are scaled to correspond to the intensity of the H89$\alpha$ line.

We studied the effects of applying the beam dilution correction to our
data as described above. By using the factors derived from 16\arcsec\
to 47\arcsec\ using the {\it Herschel}/PACS data and those from
47\arcsec\ to 84\arcsec\ and 115\arcsec\ using the WISE 22$\mu$m
observations, we find that the derived electron densities vary by
$\sim$15\% for sources observed in RRLs at 84\arcsec and by $\sim$36\%
for sources observed at 115\arcsec.

\begin{deluxetable}{lccccc} 
\tabletypesize{\footnotesize} \centering \tablecolumns{5} \small
\tablewidth{0pt} \tablecaption{Beam dilution correction}
 \tablenum{5}
\tablehead{\colhead{LOS} &  \colhead{Velocity} & \colhead{$I_{15.7\arcsec}/I_{47\arcsec}$} & \colhead{$I_{84\arcsec}/I_{47\arcsec}$} & \colhead{$I_{115\arcsec}/I_{47\arcsec}$} & \colhead{$n^*_e/n_e^1$}  \\ 
\colhead{}  & \colhead{[km\,s$^{-1}$]} &   \colhead{} &    \colhead{}  &    \colhead{} &    \colhead{}  \\}
\startdata
G305.1+0.0   &  -33.2  &  1.11  &  $-$ &   0.97  &  1.26  \\ 
G316.6+0.0   &  -48.1  &  1.11  &  $-$ &   1.26  &   $-^2$  \\ 
G316.6+0.0   &  -6.4  &  1.11   &  $-$ &   1.26  &  2.03  \\ 
G342.2+0.0   &  -131.2  &  1.18 &  $-$ &   1.00  &  1.61  \\ 
G337.0+0.0   &  -121.5  &  1.22 &  $-$ &   1.00  &  1.54  \\ 
G337.0+0.0   &  -76.6  &  1.22  &  $-$ &   1.00  &  1.31  \\ 
G345.7+0.0   &  -122.8  &  1.25 &  $-$ &   1.00  &  1.35  \\ 
G345.7+0.0   &  -8.2  &  1.25   &  $-$ &   1.00  &  1.10  \\ 
G349.1+0.0   &  17.0  &  1.14   &   1.01 & $-$   &  1.18  \\ 
G349.1+0.0   &  -91.1  &  1.14  &   1.01 & $-$  &  1.20  \\ 
G013.9+0.0   &  45.5  &  0.99  &   0.89 & $-$ &  0.83  \\ 
G013.9+0.0   &  30.3  &  0.99  &   0.89 & $-$ &  0.90  \\ 
G030.0+0.0   &  95.4  &  1.15  &   0.96 & $-$ &  1.28  \\ 
G031.3+0.0   &  100.4  &  1.07  &   0.96& $-$ &  1.18  \\ 
G031.3+0.0   &  38.0  &  1.07  &   0.96 & $-$ &  1.03  \\ 
G049.1+0.0   &  59.7  &  0.89  &   1.05 & $-$ &  0.80  \\ 
\enddata
\tablenotetext{1}{Ratio of the beam dilution corrected electron density ($n^*_e$) to that resulting when no beam dilution is applied ($n_e$).}
\tablenotetext{2}{The electron density derived without a beam dilution correction in this velocity component is undefined.}
\label{tab:dilution}
\end{deluxetable}

\section{Gaussian Decomposition}
\label{sec:gauss-decomp}

In Table\,\ref{tab:gauss-decomp-1}, we show central velocities, full
width at half maximum (FWHM), and peak intensity resulting from the
Gaussian decomposition to our RRL and [N\,{\sc ii}] 205$\mu$m
observations. Integrated intensities are listed in
Table\,\ref{tab:intensities}.

\begin{deluxetable}{lccc} 
\tabletypesize{\footnotesize} \centering \tablecolumns{5} \small
\tablewidth{0pt} \tablecaption{RRL and [N\,{\sc ii}] Gaussian Decomposition}
 \tablenum{7}
\tablehead{\colhead{LOS} &  \colhead{Velocity }  & \colhead{ FWHM}  & \colhead{$T^{\rm RRL}_{\rm mb}$}    \\ 
\colhead{}  & \colhead{[km\,s$^{-1}$]} &   \colhead{[km\,s$^{-1}$]} &    \colhead{[$10^{-3}$K]}   }
\startdata
 H91$\alpha$\\
\hline
 G305.1+0.0 & -33.2$\pm$0.2 & 23.3$\pm$0.5 & 63.6$\pm$3.6          \\
 G316.6+0.0 & -6.4$\pm$1.3 & 24.2$\pm$2.8 & 4.7$\pm$0.7            \\
 G316.6+0.0 & -48.1$\pm$0.6 & 27.6$\pm$1.5 & 19.5$\pm$1.8          \\
 G337.0+0.0 & -121.5$\pm$0.5 & 19.0$\pm$1.3 & 55.6$\pm$5.5         \\
 G337.0+0.0 & -76.6$\pm$0.8 & 28.3$\pm$2.2 & 43.3$\pm$5.5          \\
 G342.2+0.0 & -131.2$\pm$0.6 & 15.4$\pm$1.2 & 18.5$\pm$3.4         \\
 G345.7+0.0 & -122.8$\pm$0.0 & 23.2$\pm$6.8 & 14.0$\pm$1.1         \\
 G345.7+0.0 & -10.2$\pm$0.1 & 32.2$\pm$0.2 & 61.5$\pm$1.1          \\
 G349.1+0.0 & 17.1$\pm$0.3 & 20.2$\pm$0.7 & 104.5$\pm$11.4   \\
 G349.1+0.0 & -93.4$\pm$1.3 & 25.0$\pm$2.7 & 27.7$\pm$11.4   \\
\hline
 H89$\alpha$\\
\hline
 G349.1+0.0 & 17.0$\pm$0.1 & 23.2$\pm$0.1 & 131.8$\pm$2.9 \\ 
 G349.1+0.0 & -91.1$\pm$0.1 & 24.7$\pm$0.3 & 25.3$\pm$2.9 \\
 G000.0+0.0 & -60.5$^{1}$ & 29.6$\pm$0.1 & 120.9$\pm$4.0 \\
 G000.0+0.0 & -37.1$^{1}$ & 44.1$\pm$0.4 & 51.1$\pm$4.0 \\
 G000.0+0.0 & 12.8$^{1}$ & 40.3$\pm$0.5 & 31.1$\pm$4.0  \\
 G000.0+0.0 & 95.0$^{1}$ & 29.5$\pm$0.4 & 22.2$\pm$4.0  \\
 G013.9+0.0 & 45.5$^{1}$ & 21.2$\pm$0.7 & 6.1$\pm$2.3 \\
 G013.9+0.0 & 30.3$^{1}$ & 19.1$\pm$0.3 & 24.3$\pm$2.3 \\
 G030.0+0.0 & 95.4$\pm$0.9 & 23.3$\pm$0.5 & 15.3$\pm$2.5 \\
 G031.3+0.0 & 100.4$\pm$0.1 & 19.2$\pm$0.2 & 31.0$\pm$2.3 \\
 G031.3+0.0 & 38.0$\pm$0.3 & 47.7$\pm$0.7 & 10.0$\pm$2.3 \\
 G049.1+0.0 & 59.7$\pm$0.3 & 18.9$\pm$0.6 & 6.6$\pm$1.1 \\
\hline
 H67$\alpha$\\
\hline
 G345.7+0.0 & -8.2$\pm$1.2  &   23.8$\pm$2.7  & 35.05$\pm$7.3  \\ 
\hline
 H54$\alpha$\\
\hline
 G000.0+0.0 &  -62.9$\pm$1.2  &  28.8$\pm$3.3  & 24.2$\pm$5 \\
\hline
[N\,{\sc ii}] & &  & [K] \\
\hline
 G305.1+0.0 & -31.6$\pm$0.3 & 22.3$\pm$0.6 & 1.6$\pm$0.1 \\ 
 G316.6+0.0 & -46.4$\pm$0.2 & 16.6$\pm$0.4 & 1.2$\pm$0.1 \\ 
 G316.6+0.0 & -6.6$\pm$1.4 & 20.8$\pm$2.6 & 0.2$\pm$0.1 \\ 
 G337.0+0.0 & -121.8$\pm$0.1 & 9.3$\pm$0.2 & 2.5$\pm$0.1 \\ 
 G337.0+0.0 & -73.5$\pm$0.7 & 23.2$\pm$2.3 & 0.9$\pm$0.1 \\ 
 G342.2+0.0 & -130.0$\pm$0.4 & 12.8$\pm$0.9 & 0.7$\pm$0.1 \\ 
 G345.7+0.0 & -121.0$\pm$0.2 & 6.9$\pm$0.6 & 0.8$\pm$0.1 \\ 
 G345.7+0.0 & -14.9$\pm$0.3 & 15.6$\pm$0.8 & 1.0$\pm$0.1 \\ 
 G349.1+0.0 & -84.4$\pm$0.6 & 16.2$\pm$1.9 & 0.7$\pm$0.1 \\ 
 G349.1+0.0 & 14.2$\pm$0.2 & 12.4$\pm$0.5 & 1.7$\pm$0.1 \\ 
 G000.0+0.0 & -60.5$\pm$0.2 & 15.6$\pm$0.6 & 2.0$\pm$0.1 \\ 
 G000.0+0.0 & -37.1$\pm$2.4 & 58.9$\pm$3.7 & 0.7$\pm$0.1 \\ 
 G000.0+0.0 & 12.8$\pm$1.4 & 28.3$\pm$3.6 & 0.4$\pm$0.1 \\ 
 G000.0+0.0 & 50.5$\pm$1.3 & 19.2$\pm$2.9 & 0.3$\pm$0.1 \\ 
 G000.0+0.0 & 95.0$\pm$1.0 & 42.4$\pm$1.0 & 0.3$\pm$0.1 \\ 
 G013.9+0.0 & 30.2$\pm$0.6 & 6.3$\pm$1.0 & 0.2$\pm$0.1 \\ 
 G013.9+0.0 & 45.3$\pm$0.5 & 7.9$\pm$1.0 & 0.3$\pm$0.1 \\ 
 G030.0+0.0 & 96.1$\pm$0.4 & 10.4$\pm$0.9 & 0.9$\pm$0.1 \\ 
 G031.3+0.0 & 101.2$\pm$0.4 & 13.9$\pm$0.8 & 1.4$\pm$0.1 \\ 
 G031.3+0.0 & 40.0$\pm$2.9 & 18.7$\pm$6.7 & 0.1$\pm$0.1 \\ 
 G049.1+0.0 & 57.8$\pm$0.3 & 7.8$\pm$0.9 & 0.4$\pm$0.1 \\ 
\enddata
\tablenotetext{1}{The  RRL Gaussian fit was done with fixed  velocities that were derived from the Gaussian fit to the [N\,{\sc ii}] spectrum.}
\label{tab:gauss-decomp-1}
\end{deluxetable}

\bibliographystyle{aasjournal}
\bibliography{papers}


\clearpage

\end{document}